\documentclass{article}
\usepackage[utf8]{inputenc}

\usepackage{amsmath}
\usepackage{graphicx}
\usepackage[T1]{fontenc}
\usepackage{babel}
\usepackage{hyperref}
\usepackage{geometry}
\numberwithin{equation}{section}

\newcommand{\eq}{\begin{equation}}
\newcommand{\eqx}{\end{equation}}
\newcommand{\eqn}{\begin{eqnarray}}
\newcommand{\eqnx}{\end{eqnarray}}

\newcommand{\f}[2]{\frac{#1}{#2}}

\newcommand{\eps}{\varepsilon}
\newcommand{\al}{\alpha}
\newcommand{\sg}{\sigma}
\newcommand{\dl}{\delta}
\renewcommand{\th}{\theta}
\newcommand{\om}{\omega}

\newcommand{\gm}{\gamma}

\newcommand{\nn}{\mathcal{N}}

\newcommand{\calO}{\mathcal{O}}

\newcommand{\qq}{\quad\quad}
\newcommand{\qqqq}{\quad\quad\quad\quad}

\newcommand{\ct}{\tilde{c}}

\newcommand{\uvcorr}{\epsilon_{UV}}

\title{OPE coefficients and the mass-gap from the integrable scattering description of 2D CFT's}

\author{Zoltan Bajnok$^{a}$\thanks{e-mail: {\tt bajnok.zoltan@wigner.hu}},\ \  
Romuald A. Janik$^{b}$\thanks{e-mail: {\tt romuald.janik@gmail.com}} \\ \\ 
\small 
${}^a$ Wigner Research Centre for Physics \\\small
H-1525 Budapest 114, P.O.B. 49, Hungary\\\small
${}^b$ Institute of Theoretical Physics\\\small
Jagiellonian University\\\small
ul. {\L}ojasiewicza 11, 30-348 Krak{\'o}w, Poland}

\date{}

\begin{document}

\maketitle

\begin{abstract}
Many two-dimensional conformal field theories have an alternative integrable scattering description, which reproduces their spectrum of conformal weights. Taking as an example the case of the Lee-Yang nonunitary CFT and the 3-state Potts minimal model,
we derive formulas, in terms of their integrable description, for the OPE coefficients of a certain specific primary operator and two identical but otherwise essentially arbitrary operators. As a side result we also obtain a novel formula for the mass-gap relation for the integrable massive deformation of the CFT. These results are obtained through expressing the first nontrivial coefficient in the UV expansion of the energy in terms of the 
integrable CFT data, i.e. the kink and anti-kink TBA solutions.
\end{abstract}

\section{Introduction}

A conformal field theory is completely specified through the knowledge of the spectrum of its conformal dimensions and of the OPE coefficients. In two dimensions, due to the Virasoro algebra we have a thorough knowledge of both ingredients for a wide range of nontrivial theories~\cite{Belavin:1984vu}. Thus the 2D CFTs form a very interesting arena of exactly solvable quantum field theories which may be used as a theoretical laboratory for exploring various issues.

In higher number of dimensions, conformal symmetry is not powerful enough, although constraints derived from numerical bootstrap yield quite nontrivial information (see~\cite{Poland:2018epd} for a review). In recent years, remarkable progress has been made in solving the four-dimensional $\nn=4$ SYM theory due to its properties of integrability~\cite{Beisert:2010jr}. This theory is of the utmost interest due to its equivalence with string theory in $AdS_5 \times S^5$~\cite{Maldacena:1997re}, and indeed the progress in integrability involved insights coming from both sides of the AdS/CFT correspondence.
Currently the spectral problem for the $\nn=4$ SYM has been essentially solved through the so-called Quantum Spectral Curve~\cite{Gromov:2013pga,Gromov:2014caa}, a significant development on top of the Y-system formulation~\cite{Gromov:2009tv}. Huge progress has also been done in the outstanding problem of finding the OPE coefficients within the hexagon framework~\cite{Basso:2015zoa}, although there, gluing and wrapping poses significant technical difficulties and requires a ``triple'' resummation.

Looking back at the developments in solving the $\nn=4$ SYM spectral problem, the community could leverage a huge body of knowledge on finding the energy spectrum of two dimensional integrable relativistic quantum field theories on a cylinder. This was far from trivial, as the $AdS_5\times S^5$ worldsheet string theory has very peculiar idiosyncrasies, but the general concepts and structures from the known relativistic setting could serve as guiding principles.  
In the case of OPE coefficients, however, we basically lack an analogous general framework, and the methods had to be developed from scratch.

The goal of the present paper is to explore how OPE coefficients are encoded in an \emph{integrable} scattering description of 2D CFTs, which is often also given by a Y-system. On the one hand, the hope is that ultimately we may uncover some interesting way that OPE coefficients are enmeshed within the integrability structures of the CFT which may be generalizable to the $\nn=4$ SYM case. On the other hand, even if this does not happen, we believe that the problem is of interest for its own sake.

A further, more technical, motivation for considering this problem in 2D CFTs is that the integrable description of those theories neccessarily requires a resummation of wrapping corrections to all orders. Thus the understanding of potential formulas for OPE coefficients in such a regime is very interesting.

Since we are motivated by the possible generalization to $\nn=4$ SYM, we would like to formulate everything \emph{only} in terms of the integrable description of the 2D CFT (i.e. the theory which we are studying) and \emph{not} in terms of its integrable massive deformation, which is much more familiar. 

In this respect, an important result of our paper is a formula for the first nontrivial term in the UV expansion of the ground state energy expressed completely in terms of the integrable CFT data i.e. the so-called kink and anti-kink TBA solutions.

In this paper we restrict ourselves to OPE coefficients involving the deformation operator and two identical operators. We study the non-unitary Lee-Yang CFT as well as the 3-state Potts minimal model CFT. In addition, we also derive similar formulas for the mass-gap relation coefficient $\kappa$, which is of crucial interest when considering the massive integrable deformations of the CFTs, and also appears in our formula for the OPE coefficients.

The plan of this paper is as follows. In section~\ref{s.preliminaries}, we recall the integrable description of some 2D CFTs and formulate our key question. We discuss
their relation to the corresponding integrable massive deformations and recall the relation between the OPE coefficient and the UV expansion of the energy. In section~\ref{s.lyformula} we derive our formula for the Lee-Yang CFT (with a generalization to the 3-state Potts model given in Appendix~\ref{s.pottsformula}). 
We also present an extension to excited states and illustrate it on the $SL(2)$ invariant vacuum and the mass-coupling relation coefficient.
In section~\ref{s.smirnov}, we look at the same problem from the perspective of an auxillary linear dressing equation and Smirnov's formulation. The goal is here again to express the results purely in terms of the integrable description of the CFT.
We close the paper with a discussion and a number of appendices containing technical details.

\section{Preliminaries}
\label{s.preliminaries}

Let us take an Euclidean CFT on the cylinder of circumference ${\cal R}$. Dimensionful quantities on the cylinder can be written in terms of dimensionless quantities on the plane. Energies depend on the volume as
\eq
E_{\calO}^{\mathrm{CFT}}({\cal R})=\frac{2\pi}{{\cal R}}\left[h_{\calO}+\bar{h}_{\calO}-\frac{c}{12}\right]
\label{Ecyl}
\eqx
where $c$ is the central charge and $h_{\calO}$ and $\bar{h}_{\calO}$ are
the left/right conformal dimensions of the primary state corresponding to the operator $\calO$.
Expectation values can be expressed in terms of CFT 3-point functions as
\eq
\langle  { \cal O}_1 \vert {\cal O} \vert { \cal O}_2 \rangle= \left( \frac{2\pi}{{\cal R}} \right)^{h_{\cal O}+{\bar h}_{\calO} }C_{ {\cal O}_1 {\cal O} { \cal O}_2}
\label{1ptcyl}
\eqx
We would like to describe the quantities, $h, \bar h $ and $ C$ in terms of an integrable formulation.

\subsection{Integrable description of 2D CFT's -- the Lee-Yang model}
\label{s.ly}

\begin{figure}
\begin{centering}
\includegraphics[width=6cm]{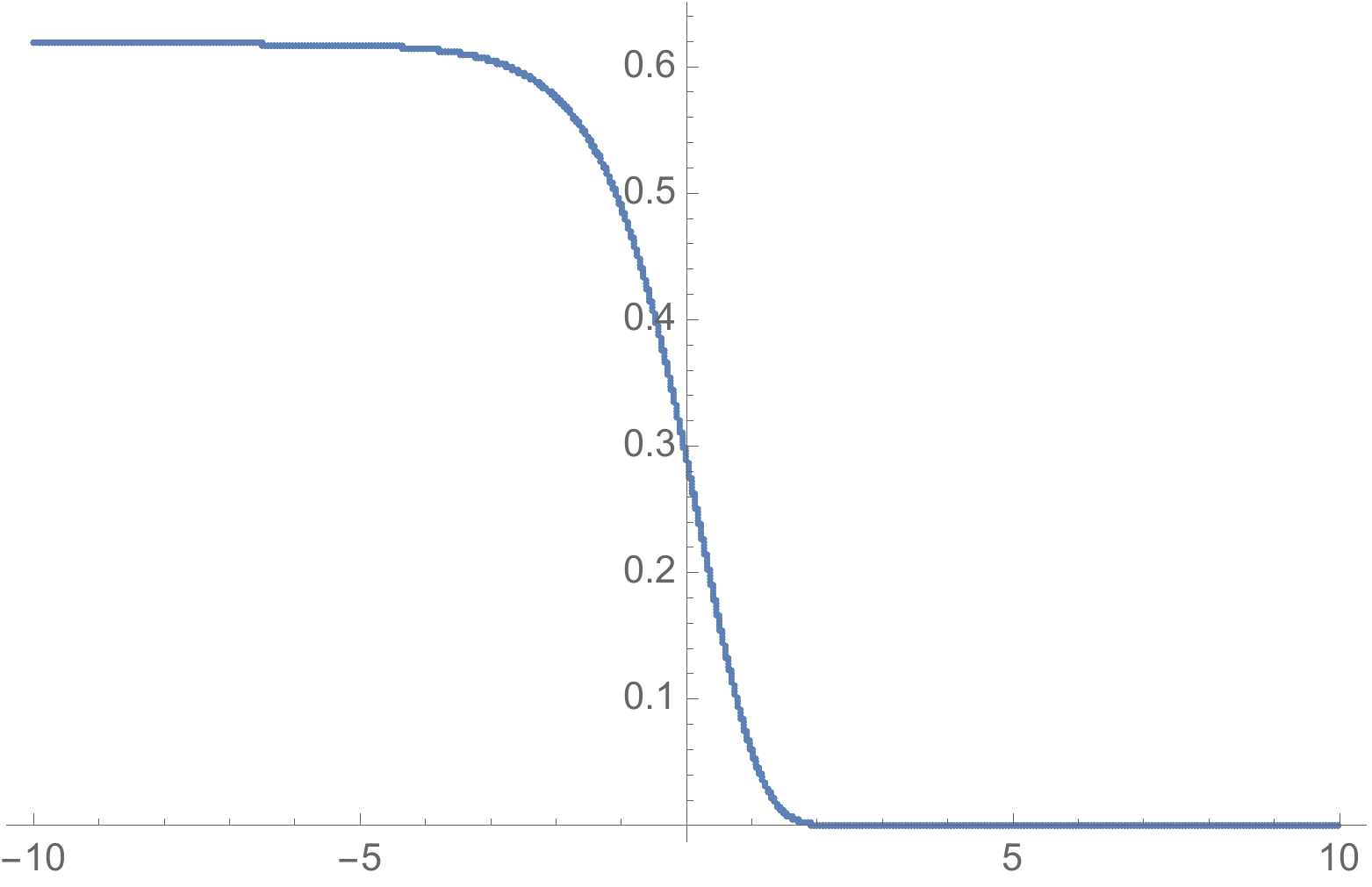}
\par\end{centering}
\caption{$Y_{R}^{-1}(s)$ as the function of $s$. }
\label{kink}
\end{figure}

Let us illustrate how the integrable formulation of 2D CFT's works for the spectrum in the simplest setting of the Lee-Yang nonunitary CFT with $c=-\f{22}{5}$. 
In this nonunitary CFT, the ground state on a cylinder corresponds to the primary state with conformal weights $(-1/5,-1/5)$. The right- and left- conformal weights can be extracted from separate Y-systems corresponding to the right-moving and left-moving sectors\footnote{We use $s$ for the rapidity in the CFT description with no volume dependence. We will use $\th$ for the rapidities in the massive integrable deformation of the CFT. 
The relations between these parameters depend on the sector.
In the right moving sector $\th = s - \log\f{r}{2}$, while in the left moving sector $\th = s + \log\f{r}{2}$.} \cite{Bazhanov:1996aq,Bajnok:2014fca}
\eq
\label{e.ysystem}
Y_{R,L}(s +i\pi/3) Y_{R,L}(s -i\pi/3) = 1 + Y_{R,L}(s)
\eqx
with the asymptotic conditions
\eq
Y_R(s) \sim_{s \to +\infty} \exp(e^s) \qqqq
Y_L(s) \sim_{s \to -\infty} \exp(e^{-s})
\label{e.ysystemasymp}
\eqx
Writing $Y_{R,L}(s) = e^{\eps_{R,L}(s)}$ this can be translated into TBA equations for the kink/anti-kink:
\eq
\label{e.cftsolutions}
\eps_{R,L}(s) = e^{\pm s} -  \phi \star \log\left(1+ e^{-\eps_{R,L}}\right)
\eqx
with the upper sign corresponding to the kink (right moving sector), see Fig.~\ref{kink}, and the lower to the antikink (left moving sector).
Note that there is no volume dependence in (\ref{e.cftsolutions}), in accordance with the conformal symmetry of the theory. Modifying the exponential source terms like $e^{\pm s} \to \f{r}{2} e^{\pm s}$ can be always reabsorbed by a translation in $s$ and would give the trivial $1/{\cal R}$ volume dependence of the energy levels on the cylinder.

The dimensionless energy $E_{\calO}^{\mathrm{CFT}}({\cal R}){\cal R}$ containing the conformal dimensions of the ground state together with the central charge  is given by the sum of the contributions of the two sectors
\eq
E_{\calO}^{\mathrm{CFT}}({\cal R}){\cal R} = 2\pi \left(h_0+\bar{h}_0-\frac{c}{12}\right) = E_R  + E_L
\eqx
where
\eq
E_R= -\int_{-\infty}^\infty \f{ds}{2\pi}  e^s  \log\left(1+ e^{-\eps_R}\right) \qq
E_L= -\int_{-\infty}^\infty \f{ds}{2\pi}  e^{-s}  \log\left(1+ e^{-\eps_L}\right)
\eqx
The outcome of this computation, which can done analytically using the so-called dilogarithm trick, of course reproduces the well known Lee-Yang answer \cite{Zamolodchikov:1989cf}
\eq
E_R+E_L = -\frac{2\pi}{30}
\eqx
In the following we will need some more detailed properties of the CFT solutions $Y_{R,L}(s)$ shown on Figure \ref{kink}.
The nonzero constant asymptotic values of the two solutions is given by a constant solution of the Y-system (\ref{e.ysystem}), which for the ground state, corresponding to the $(h_0, \bar{h}_0) = (-1/5,-1/5)$ primary, is
\eq
Y_0 = \f{1}{2}\left( 1 + \sqrt{5} \right)
\eqx
The precise way of how the CFT kink/antikink approach this constant value will be of crucial importance for our further considerations. There is a systematic expansion around the asymptotic value in powers of $e^{\pm(1-h)s}$, where $h=-1/5$:
\eq
Y_{R,L}^{-1}(s) Y_0 = 1 - \ct\, \underbrace{e^{\pm(1-h)s}}_{e^{\pm \f{6}{5}s}} + \ldots
\qq \text{as}\quad s \to \mp \infty
\label{e.tail}
\eqx
where the upper signs correspond to the right-moving (kink-like) TBA solution. The expansion coefficients, in particular $\tilde c$ are related to the non-local integrals of motions. 
In general, the coefficient $\ct$ can be determined numerically, but for the Lee-Yang CFT there exists an analytical expression which follows from the formulas derived\footnote{Note, however, that even in~\cite{Bazhanov:1996aq} this coefficient has not been determined purely from TBA considerations.} in~\cite{Bazhanov:1996aq}.
\eq
\tilde c=\pi^{-\frac{11}{10}}\left(\sqrt{5}-5\right)\Gamma\left(-\frac{4}{5}\right)\Gamma\left(\frac{11}{10}\right)\left(\frac{\Gamma\left(\frac{2}{3}\right)}{\Gamma\left(\frac{1}{6}\right)}\right)^{\frac{6}{5}}=0.785438261\dots 
\eqx

The other primary field in the Lee-Yang CFT is the $SL(2)$ invariant vacuum with conformal weights $(h_1, \bar{h}_1) = (0,0)$. In the integrable description, it is given again by solutions of the same Y-system (\ref{e.ysystem})-(\ref{e.ysystemasymp}) but with different analyticity conditions, which can be equivalently formulated either as a zero of $Y_{R,L}$ on the real axis at a certain $s=\alpha$~\cite{Bazhanov:1996aq} 
\eq
Y_{R,L}(\alpha) = 0
\eqx
or as a pair of ``singularities'' at $s=\alpha \pm i\pi/3$~\cite{Dorey:1996re}
\eq
Y_{R,L}\left(\al \pm i \f{\pi}{3}\right)= -1
\eqx
The asymptotic value of the kink/anti-kink solutions is now given by the second solution of the constant Y-system
\eq
Y_0 = \f{1}{2}\left( 1 - \sqrt{5} \right)
\eqx
and the approach to this asymptotic value has a different exponential scaling
\eq
Y_{R,L}^{-1}(s) Y_0 = 1 - \ct'\, e^{\pm\f{12}{5}s} + \ldots
\qq \text{as}\quad s \to \mp \infty
\eqx
In fact, the coefficient $\ct'$ is related to the \emph{second} nonlocal conserved charge, as the first nonlocal conserved charge in this case vanishes.
This different exponential scaling will be important for our considerations. 
Let us note that the specific power $\gm$ of the exponent in
\eq
Y_R^{-1} Y_0 =  1 - \ct\, e^{\gm s} + \ldots
\eqx
can be determined by substitution into the Y-system (\ref{e.ysystem}) and depends on the asymptotic value $Y_0$. We get the relation
\eq
\cos \f{\pi}{3} \gm = (2 Y_0)^{-1}
\eqx
which reproduces $\gm=\f{6}{5}$ for the $(-1/5, -1/5)$ primary state and $\gm=\f{12}{5}$ for the $(0, 0)$ $SL(2)$ invariant vacuum.

\subsection{The key question}

The key question that we would like to investigate is to find how to express the remaining fundamental CFT quantities -- the OPE coefficients directly in terms of the integrable data defining the relevant states like $Y_{R,L}$ given above. 

In this paper we would like to concentrate on the simplest case of symmetric OPE coefficients of the form
\eq
C_{\calO \Phi \calO}
\label{e.opesimple}
\eqx
where $\Phi$  is a CFT operator which generates a massive integrable deformation with a massive TBA description. 
The OPE coefficients are also related to 1-point functions on a cylinder \eqref{1ptcyl}. 
In the specific case of $C_{\calO \Phi \calO}$,
these expectation values can in turn be related to the expansion of the energy of the state $\calO$ at small volume in the massive integrable deformation of the CFT by the operator $\Phi$, which we review below.

\subsection{Massive integrable deformations}

Consider perturbing the CFT with a spinless relevant primary
field of dimension $(h,\bar{h}=h)\footnote{Operators normalized canonically 
 in the CFT: $\Phi(z,\bar z)\vert \Phi(0,0)= z^{-2h}{\bar z}^{-2\bar h}+\dots $.}$: 
\eq
S=S_{\mathrm{CFT}}+\lambda\int_{-\infty}^{\infty}dy\int_{0}^{\cal R}dx\,\Phi(x,y)
\eqx
such a way that the theory remains integrable, i.e. there remains infinitely
many higher spin conserved charges. For generic minimal models there
are many inequivalent ways to do this, however for the Lee-Yang model the only choice
is $h=-\frac{1}{5}$.

\subsubsection{UV energy expansion and the OPE coefficient}

The change in the energy levels of the massive deformation of the CFT can be calculated perturbatively
\eq
E_{\calO}({\cal R})=E_{\calO}^{\mathrm{CFT}}({\cal R})+\frac{2\pi}{{\cal R}}\left[\sum_{n=1}^{\infty}c_{n}\lambda^{n}\left(\frac{{\cal R}}{2\pi}\right)^{n(2-2h)}\right]
\label{e.cftuv}
\eqx
where $c_{1}$ is related to the expectation value of the operator
as 
\eq
c_{1}=2\pi C_{\calO\Phi\calO}
\eqx
while higher $c_{n}$-s are related to integrated
multipont correlation functions.

For the $SL(2)$ invariant CFT vacuum state, $c_{1}$ vanishes and the
leading correction comes from integrating the universal two-point
function, which results  in a model-independent expression:
\begin{equation}
c_{2}(h)=2\pi \int_{\vert z\vert<1}d^{2}z(z\bar{z})^{h-1}\langle0\vert\Phi(1,1)\Phi(z,\bar{z})\vert0\rangle=\pi^2\frac{\Gamma(h)^{2}\Gamma(1-2h)}{\Gamma(1-h)^{2}\Gamma(2h)}
\label{c2h}
\end{equation}

For the integrable deformation of the non-unitary Lee-Yang CFT discussed above (the Scaling Lee-Yang Model), the resulting expansion takes the following forms for the states corresponding to the two primary fields of the CFT. 

The lowest energy
state corresponds to $\Phi$ and the leading energy correction comes
from $c_{1}=2\pi C_{\Phi\Phi\Phi}$ in the form
\eq
E_{\Phi}({\cal R})=\frac{2\pi}{{\cal R}}\left(-\frac{1}{30}+2\pi\lambda C_{\Phi\Phi\Phi}\left(\frac{{\cal R}}{2\pi}\right)^{\frac{12}{5}}+\dots\right)
 \label{e.cftenergycorrphi}
\eqx

The first excited state at small volumes  corresponds to the $SL(2)$ invariant CFT state, whose leading correction goes as 
\eq
E_{I}({\cal R})=\frac{2\pi}{{\cal R}}\left(\frac{11}{30}+\lambda^2 c_2(-1/5)\left(\frac{{\cal R}}{2\pi}\right)^{\frac{24}{5}}+\dots\right)
 \label{e.cftenergycorrid}
\eqx
The expansion of the energy levels goes in powers of the dimensionful coupling constant $\lambda $, which sets the scale in the massive scattering theory. 

\subsubsection{UV energy expansion from TBA}

As the massive deformation of the CFT was chosen to be integrable, the energy of the states can also be computed from the Thermodynamic Bethe Ansatz (TBA), which we review below. We also identify the precise way to extract the OPE coefficient of interest from the TBA expansion, taking into account the mass-coupling relation.

When the perturbation is integrable, the scattering matrix, which connects
infinite volume asymptotic states, factorizes into two-particle scatterings.
In the simplest case of one single particle type of mass $m$, the two-particle S-matrix
is a simple function of the rapidity differences $S(\theta)$, which
satisfies unitarity $S(\theta)S(-\theta)=1$ and crossing symmetry
$S(i\pi-\theta)=S(\theta)$. In the Scaling Lee-Yang model it takes the form
\cite{Cardy:1989fw}
\eq
S(\theta)=\frac{\sinh\theta+i\sin(\pi/3)}{\sinh\theta-i\sin(\pi/3)}
\eqx
The pole at $\theta=2\pi i/3$ signals a boundstate, which is the
particle itself. 
This S-matrix can be used to calculate the finite volume groundstate\footnote{For operators $\calO$ corresponding to excited states, one has to add additional terms as described below.}
energy \cite{Zamolodchikov:1989cf}
\eq
E^{TBA}_\calO(r)=-m\int\frac{d\theta}{2\pi}\cosh\theta\,\log(1+e^{-\eps(\theta)})
\eqx
where the pseudo-energy satisfies the TBA equation 
\eq
\label{e.tbamassive}
\eps(\theta)=r\cosh\theta-\int\frac{d\theta'}{2\pi}\phi(\theta-\theta')\log(1+e^{-\eps(\theta')})=r \cosh \theta - \phi \star \log(1+e^{-\varepsilon})
\eqx
with $\phi(\theta)$ being the logarithmic derivative of the scattering
matrix $\phi(\theta)=-i\partial_{\theta}\log S(\theta)$ and $r=m{\cal R}$
is the dimensionless volume. This formula can be extended for excited
states by analytical continuation \cite{Dorey:1996re,Dorey:1997rb}, which results in source
terms in the TBA equation. These can be constant twists, or sums of
the logarithms of S-matrices. The same equations could be derived from the integrable structure via functional relations \cite{Bazhanov:1996aq} or from the continuum limit of an integrable lattice regularization \cite{Bajnok:2014fca}. 

Let us make a few remarks here.  First, the TBA energy is written in terms of the mass of the scattering particles, while the perturbed CFT energy is expressed in terms of the coupling constant $\lambda$. On dimensional grounds the mass and the coupling can be related as 
\eq
\lambda = \kappa m^{2-2h}
\label{e.massgapdef}
\eqx
where $\kappa$ is a very crucial  model-dependent constant\footnote{In multiparameter deformations it can be a non-trivial function of the dimensionless coupling \cite{Bajnok:2015eng,Bajnok:2016ocb}.}, which connects the UV and IR descriptions. It is in fact quite nontrivial --
in the Lee-Yang theory its value is \cite{Zamolodchikov:1995xk}
\eq
\kappa=\frac{5^{\frac{3}{4}} \left(\Gamma \left(\frac{2}{3}\right) \Gamma
   \left(\frac{5}{6}\right)\right)^{\frac{12}{5}}}{8\ 2^{\frac{4}{5}} \pi ^{\frac{6}{5}} \Gamma
   \left(\frac{3}{5}\right) \Gamma \left(\frac{4}{5}\right)}
\eqx
We will show shortly how it can be extracted from our approach. Formula (\ref{e.massgapdef}) is also called the mass-gap relation as it expresses the physical mass in terms of the Lagrangian coupling.

Second, the TBA groundstate energy $E^{TBA}_\calO(r)$ and the perturbed CFT energy $E_{\calO}({\cal R}) $ are given in different schemes: $E^{TBA}_\calO(r)$ is chosen to vanish for large volumes, while $E_{\calO}({\cal R})$ has a nonzero bulk energy density $\epsilon_B$.  They are related to each other as 
\eq
E_{\calO}({\cal R})=\epsilon_B {\cal R}+ E^{TBA}_\calO(r) \qqqq \epsilon_B= -\frac{m^2}{4\sqrt{3}} \quad \text{(for SLYM)}
\label{ECPTvsETBA}
\eqx

A key quantity of interest for the present paper is the first nontrivial term $\uvcorr$ in the UV expansion of the TBA energy
\eq
{\cal R} E^{TBA}_\calO(r) = 2 \pi \left( h_\calO+\bar{h}_\calO - \f{c}{12} \right)
-\frac{\epsilon_B}{m^2} r^2 + \uvcorr r^\al + \ldots
\label{e.defuvcorr}
\eqx
In the TBA scheme, the bulk energy appears in the small volume expansion of the energy with an opposite sign and called, for this reason, the anti-bulk energy. 

Depending on the TBA state in question, the numerical coefficient $\uvcorr$ can be related either to the OPE coefficient $C_{\calO \Phi \calO}$ through
\eq
C_{\calO \Phi \calO} = \f{1}{\kappa (2\pi)^{2h}} \cdot \uvcorr
\label{e.cftenergycorr}
\eqx
when $\alpha=2-2h$ in (\ref{e.defuvcorr}), or to the mass-coupling relation $\kappa$:
\eq
\kappa^2 = \f{ (2\pi)^{2(2-2h)}}{2\pi c_2(h)} \cdot \uvcorr
\label{e.kappasq}
\eqx
when the state is the $SL(2)$ invariant vacuum $\calO \equiv 1$ (in this case $\alpha=2(2-2h)$). 
We will provide a ``master formula'' for $\uvcorr$ purely in terms of the CFT integrable data. Consequently, this will provide formulas for the above OPE coefficient and the mass-coupling relation $\kappa$. 




\section{A formula for the UV energy expansion, the OPE coefficient and the mass-gap}
\label{s.lyformula}

As we reviewed in the previous section, the OPE coefficient of the deforming operator and the mass-coupling relation can be read off from the first nontrivial coefficient $\uvcorr$ of the small volume expansion of the energy.
In this section we will derive our main result, a formula for this coefficient expressed purely in terms of the integrable CFT data described in section~\ref{s.ly}.
The \emph{analytical} determination of this coefficient directly from TBA remains in fact an outstanding problem. We hope that our formula may be a useful step in this direction, but the possibility of an analytical evaluation is not really that important for our purposes.

 Here we perform our calculations for the simplest case of the Scaling Lee-Yang model. In Appendix~\ref{s.pottsformula}, we show that a similar formula holds in the two-component TBA of the (scaling) 3-state Potts model\footnote{In the 3-state Potts model we consider the nontrivial $\sg$ state instead of the vacuum, which thus does not trivially reduce to the Scaling Lee-Yang model TBA. It is also this state which is interesting from the point of view of OPE coefficients.}.

In the following, instead of considering the TBA $\eps(\th)$, it is more convenient to employ $Y(\th)=e^{\eps(\th)}$ (or more precisely $Y^{-1}(\th) = e^{-\eps(\th)}$, which has better behaviour on the real line).
The TBA equation of the massive integrable deformation has the form
\eq
\log Y = \f{r}{2} e^\th + \f{r}{2} e^{-\th} - \phi \star \log\left( 1 +Y^{-1} \right)
\eqx
The asymptotic solution for small $r$ can be written as
\eq
\label{e.Yas}
Y_{\mathrm {as}} = Y_K Y_A Y_0^{-1}
\eqx
where $Y_{K,A}$ are the standard kink and antikink solutions defined by
\eq
\label{e.YKA}
\log Y_{K,A} = \f{r}{2} e^{\pm \th} - \phi \star \log\left( 1 +Y_{K,A}^{-1} \right)
\eqx
Both of them can be expressed in terms of the volume independent CFT solutions (\ref{e.cftsolutions}) as
\eq
\label{e.kinkRL}
Y_K(\th) \equiv Y_R\left(\th + \log \f{r}{2}\right) \qqqq
Y_A(\th) \equiv Y_L\left(\th - \log \f{r}{2}\right)
\eqx
The kink and antikink thus inherit the asymptotics (\ref{e.tail}).
As we will see shortly, it is exactly this exponential tail of the kink in the region of the antikink (and vice versa) which generates a correction of order $r^{\f{12}{5}}$ which is responsible for the energy correction (\ref{e.cftenergycorr}) that we are after.

\begin{figure}
\begin{centering}
\includegraphics[width=8cm]{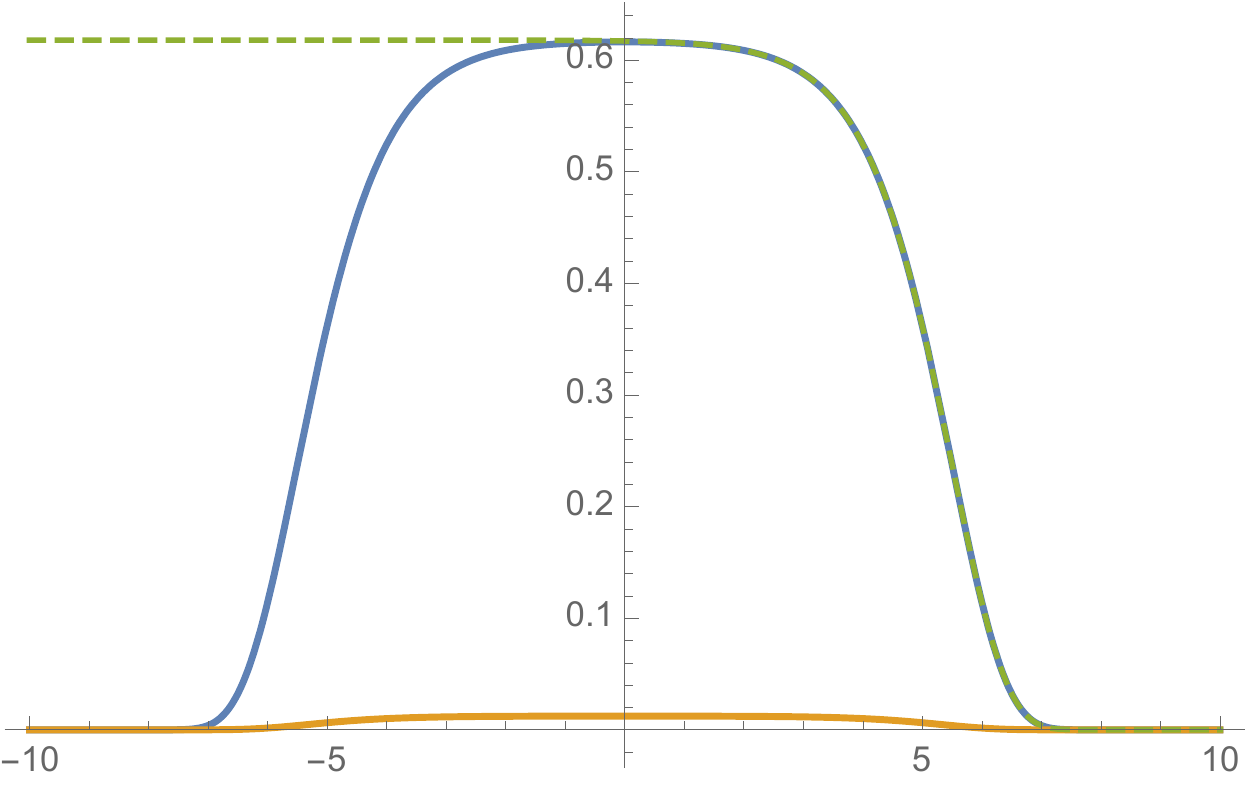}
\par\end{centering}
\caption{$Y^{-1}(\theta)$ as the function of $\theta$ for dimensionless volume $r=0.01$ in blue. Dashed green shows the kink solutions. The difference between $Y^{-1}-Y^{-1}_{\mathrm{as}}=\delta Y^{-1}$ is of order $10^{-7}$ and is not visible. Thus for demonstration, we draw $r^{-\frac{12}{5}} \delta Y^{-1}$  with yellow. }

\end{figure}


\subsection{Subleading TBA solution}

We would now like to derive an equation for the first subleading correction to $Y_{\mathrm {as}}^{-1}$ which is of order $r^{2-2h}$, $r^{\f{12}{5}}$ in the Lee-Yang model. Inserting 
\eq
Y^{-1} = Y_{\mathrm {as}}^{-1} + \dl Y^{-1}
\eqx
into the TBA equation we obtain a linear integral equation for $\dl Y^{-1}$:
\eq
- Y_{\mathrm {as}} \dl Y^{-1} = \underbrace{ \f{r}{2} e^\th + \f{r}{2}e^{-\th} - \log Y_{\mathrm {as}} 
-\phi \star \log(1+Y_{\mathrm {as}}^{-1})}_{\mathrm {source}} - \phi \star \f{\dl Y^{-1}}{1+Y_{\mathrm {as}}^{-1}}
\eqx
Using the definition (\ref{e.Yas}) and the kink/antikink equations (\ref{e.YKA}) we can rewrite the source term in the following form
\eqn
\label{e.source}
\mathrm {source} &=& -\phi\star \left[ \log(1+Y_{\mathrm {as}}^{-1}) -\log(1+Y_K^{-1}) - \log(1+Y_A^{-1}) +\log Y_0 \right] \nonumber\\
&\equiv& - \phi \star G
\eqnx
We need now to extract the leading small $r$ behaviour of the source $G$, focusing separately on the two asymptotic (kink/antikink) regions of $\th$.
Moreover, we aim to rewrite the source directly in terms of the volume-independent CFT solutions $Y_{R,L}$. We want, therefore, to calculate the following quantities which are concentrated in the right moving and left moving regions
\eq
\label{e.rescaledsource}
G_{R,L}(s) = \lim_{r \to 0} r^{2h-2} G\left(s \mp \log \f{r}{2}\right)
\eqx
Let us consider $G_R(s)$ i.e. the source in the kink asymptotic region. In this region it is convenient to rewrite $G$ as
\eq
G = \log \f{1+ Y_K^{-1} Y_A^{-1} Y_0}{1+Y_K^{-1}}  - \log \f{1+Y_0^{-1} Y_A^{-1} Y_0}{1+Y_0^{-1}}
\eqx
Now using (\ref{e.kinkRL}) and (\ref{e.tail}) we see the appearance of the $r^{2-2h}$ scaling
\eq
\label{e.antikinkr}
Y_A^{-1} Y_0 -1 \sim  - \ct e^{(h-1)\left( s- 2\log \f{r}{2} \right)} = - \ct \left( \f{r}{2}\right)^{2-2h} e^{(h-1) s}
\eqx
We can simply denoted the tail of the anti-kink solution in the kink region as 
\eq
y_L(s)=\lim_{r\to 0}r^{2h-2} (Y_A^{-1}(s-\log \frac{r}{2})Y_0-1)=
-\ct\,2^{2h-2} e^{(h-1) s}
\eqx
such that the rescaled source in the kink region takes the form
\eq
\label{e.GKAR}
G_R = \left( \f{1}{1+Y_R} - \f{1}{1+Y_0} \right) y_L 
\eqx
This source behaves nicely in the $s\to +\infty$ limit, but seems at first glance dangerous for large negative $s$ (in the above formula $h=-1/5$). Fortunately, there the term in parenthesis vanishes exponentially with just the appropriate asymptotics to cancel this growth and 
the $s\to -\infty$ limit is a constant:
\eq
\lim_{s \to -\infty} G_R(s) = \f{Y_0}{(1+Y_0)^2} \, \ct^2\, 2^{2-2h} 
\eqx
The behaviour of $G_L(s)$ is clearly analogous.

We can now write the linearized equation for the leading $\dl Y^{-1}$ correction separately in the kink (right-moving) sector and antikink (left-moving) sector. Extracting further the overall $r^{2-2h}$ scaling as above we obtain
\eq
\label{e.linearizedR}
Y_R\, \dl Y_R^{-1} = \phi \star \left[ 
G_R + \f{\dl Y_R^{-1}}{1 + Y_R^{-1}}
\right]
\eqx
where the subscript on $\dl Y_R^{-1}$ denotes the appropriate sector. All quantities in (\ref{e.linearizedR}) are expressed in terms of the CFT rapidity $s$.

It is important to emphasize, that the source term in the kink sector $G_R$ involves not only the right-moving kink solution, but includes also the asymptotic tail of the left-moving anti-kink solution. Hence it involves in a nontrivial way the interplay between both sectors of the theory. 

\subsection{Evaluation of the UV energy correction coefficient $\uvcorr$}

In order to compute the energy $ {\cal R} E^{TBA}_\calO(r)$, we need to evaluate the integral
\eq
-r\int \f{d\th}{2\pi} \cosh \th  \log\left( 1+ Y_{\mathrm {as}}^{-1} + \dl Y^{-1} \right)=  
\underbrace{E_R+E_L}_{2\pi \left(h_0+\bar{h}_0-\frac{c}{12}\right)}
-\frac{\epsilon_B}{m^2}r^2+\uvcorr r^{2-2h}+\dots
\eqx
where in the Lee-Yang model $h_0=h=-\frac{1}{5}$ for the ground state.  We are interested in the first nontrivial $r\to 0$ contribution of order $r^{2-2h}$. Expanding the above integral we get two contributions
\eq
-r\int \f{d\th}{2\pi} \cosh \th  \log\left( 1+ Y_{\mathrm {as}}^{-1} \right) 
-r\int \f{d\th}{2\pi} \cosh \th \f{\dl Y^{-1}}{1+ Y_{\mathrm {as}}^{-1}}
\label{e.twocontrib}
\eqx
The second term will already contribute at order $r^{2-2h}$. To see this we split the $\cosh\th$ into exponentials and integrate $e^\th$ in the kink region (and $e^{-\th}$ in the anti-kink region):
\eq
-\int \f{d\th}{2\pi} \f{r}{2} e^\th \f{\dl Y^{-1}}{1+ Y_{\mathrm {as}}^{-1}}
\quad\longrightarrow\quad
- \int \f{ds}{2\pi} e^s  
\f{\dl Y_R^{-1} \cdot r^{2-2h}}{1+ Y_R^{-1}}
\eqx
Recall that we factored out $r^{2-2h}$ when defining $\dl Y_R^{-1}$.

We need to be careful as a contribution of the same order will also arise from the first term in (\ref{e.twocontrib}), where the tail of the anti-kink appearing in $Y_{\mathrm {as}}$ in the kink region will lead to this scaling similarly to (\ref{e.antikinkr}). In order to isolate this contribution, it is convenient to use the definition of the source $G$ from (\ref{e.source}) and rewrite
\eq
\log\left( 1+ Y_{\mathrm {as}}^{-1} \right) = \log(1+Y_K^{-1}) + \log(1+Y_A^{-1}) - \log(1+Y_0^{-1}) + G
\label{e.Yassum}
\eqx
In order to isolate the $r$ scaling, we have to again split the $r\cosh \th$ into exponentials and take into account the resulting powers of $r$ arising from moving to the kink or anti-kink regions.
E.g. plugging the first term in (\ref{e.Yassum}) into the energy formula with $\f{r}{2}e^{\th}$ and evaluating it in the kink region 
\eq
-\int \f{d\th}{2\pi} \f{r}{2}e^{ \th   }\log\left( 1+ Y_{\mathrm {K}}^{-1} \right) =
-\int \f{ds}{2\pi} e^{ s  }\log\left( 1+ Y_{\mathrm {R}}^{-1} \right) 
\eqx
we obtain the right conformal energy, which can be calculated using the dilogarithm trick. 

In the second and third term, again with the exponential $\f{r}{2}e^{\th}$, we can make an integration by parts, together with a shift in the integral into its antikink domain
\eq
-\left(\frac{r}{2}\right)^2\int \f{ds}{2\pi} e^{ s  }\frac{\partial \log Y_{L}}{1+Y_L} 
\eqx
This integral gives the anti bulk energy constant in \eqref{ECPTvsETBA}, which can be evaluated by inspecting the asymptotics of the derivative of the anti-kink equation. 

Finally, $G$ will give a contribution at the desired subleading order $r^{2-2h}$, which we have to evaluate together with the contribution coming from $\delta Y^{-1}$. In doing so 
we can again split the subleading correction into the kink and antikink regions, which can be computed in a similar manner. We will now focus on the kink regime. 
We thus obtain
\eq
\label{e.subleadingtbaE}
-\int \f{ds}{2\pi} e^s \left[ G_R + \f{\dl Y_R^{-1}}{1+Y_R^{-1}} \right]
\eqx
Note that the term in brackets is exactly the combination appearing in the linearized equation for $\dl Y_R^{-1}$ in the kink region (\ref{e.linearizedR}).
Fortunately, we do not need to find $\dl Y_R^{-1}$ explicitly, but we can use an analog of the dilogarithm trick to evaluate (\ref{e.subleadingtbaE}) purely in terms of the properties of the asymptotic CFT solutions $Y_{R,L}$. The details of this computation are given in Appendix~\ref{s.dilogenergy} and the resulting formula involves just the UV limit CFT solutions $Y_{R,L}$ of section~\ref{s.ly}.
\eq
-\int \f{ds}{2\pi}  G_R \cdot \partial \log Y_R
\eqx
The source term $G_R$ given by (\ref{e.GKAR}) involves the right moving solution $Y_R$ together with the \emph{exponential tail} of the left moving one $Y_L$. We will find that this overall structure is in fact quite generic.
Adding an analogous contribution from the antikink side of the energy integral, we get our final formula for the key energy correction coefficient $\uvcorr$:
\eq
\label{e.deltaEly}
 \uvcorr = -\int \f{ds}{2\pi}  G_R \cdot \partial \log Y_R
- \int \f{ds}{2\pi} G_L \cdot \partial \log Y_L
\eqx

\subsection{Formula for the OPE coefficient}

Using the relation (\ref{e.cftenergycorr}) from section~\ref{s.preliminaries}, we obtain the formula for the OPE coefficient
\eq
\label{e.opely}
C_{\Phi\Phi\Phi} = -\frac{(2\pi)^{\frac{2}{5}}}{\kappa} \cdot \left[ \int \f{ds}{2\pi}  G_R \cdot \partial \log Y_R
+ \int \f{ds}{2\pi} G_L \cdot \partial \log Y_L \right]
\eqx
Although we cannot evaluate the above formula analytically, we can compute it very precisely numerically, which is close to the exact answer
\eq
C_{\Phi\Phi\Phi} =1.91131...=\sqrt{\frac{2}{1+\sqrt{5}}}\frac{ \Gamma \left(\frac{1}{5}\right) \Gamma \left(\frac{6}{5}\right)}{\Gamma
   \left(\frac{3}{5}\right) \Gamma \left(\frac{4}{5}\right)}
\eqx

In Appendix~\ref{s.pottsformula} we show that a direct analog of the above formulas also exists for the 3-state Potts model, where the energy correction is related to the $C_{\sg \eps\sg}$ OPE coefficient.

Several comments are in order here. First, it is encouraging that the OPE coefficient can be expressed purely in terms of the integrable CFT description without the need to explicitly go away from the UV point using the massive integral deformation.
Indeed, the integrand in (\ref{e.opely}) involves quantities defined directly through the Y-system of the CFT.
Second, the expression involves an interplay between the right-moving and left-moving sector through the tail of the antikink appearing in $G_R$.
Hence it is a much more nontrivial quantity than the energy levels (conformal weights).
Third, the link of the energy formula with the OPE coefficient involves a nontrivial normalization factor incorporating the mass-gap coefficient $\kappa$. We will show in the next subsection, however, that this coefficient can also be expressed in terms of the CFT integrable data, but for the $SL(2)$-invariant vacuum state.

Finally, let us discuss the freedom of changing the operators in the OPE coefficient. In the Lee-Yang CFT, an analogous computation could be done starting with the TBA solution of the identity operator with $(h_1, \bar{h}_1)=(0,0)$. Clearly
\eq
C_{1\Phi 1} = 0
\eqx
We would like now to understand how this vanishing is recovered in the integrable framework.
The difference lies in the asymptotics of the right- and left-moving CFT TBA solutions. Apart from an inessential change of the asymptotic value which is now equal to $Y_1 = (1-\sqrt{5})/2$, the crucial modification is the change in the exponential approach to the constant value.
Indeed, in contrast to (\ref{e.tail}) we have
\eq
Y_{R,L}^{-1}(s) Y_1 = 1 - \ct'\, e^{\pm\f{12}{5}s} + \ldots
\qq \text{as}\quad s \to \mp \infty
\eqx
so the term with $e^{\pm\f{6}{5}s}$ is absent,
which is associated to the vanishing of the first nonlocal conserved charge in the language of~\cite{Bazhanov:1996aq}.
This implies that due to the limit in (\ref{e.rescaledsource}), the source terms $G_{R,L}$ vanish and we recover $C_{1\Phi 1} = 0$. 

Note, however, that if we extract instead $r^{2(2-2h)} = r^{\f{24}{5}}$ from the source term, the computation of the energy correction (\ref{e.deltaEly}) goes through. The result should not be interpreted as an OPE coefficient, but rather it provides the mass-coupling relation coefficient $\kappa$. Let us now investigate this state in detail.

\subsection{The $SL(2)$ invariant excited state and the mass-coupling formula}

The excited state TBA equation for a one-component system has extra
source terms of the form 
\begin{equation}
\log Y=r\cosh\theta+\sum_{i}\eta_{i}\log S(\theta-\theta_{i})-\phi\star\log(1+Y^{-1})
\end{equation}
where the roots $\theta_{i}$ are located at the zeros of the logarithm
\begin{equation}
Y(\theta_{i})=-1
\end{equation}
The energy also gets correction from the roots as
\begin{equation}
E_I^{TBA}=-im\sum_{i}\eta_{i}\sinh\theta_{i}-m\int\frac{d\theta}{2\pi}\cosh\theta\log(1+Y^{-1})\label{eq:exELY}
\end{equation}
These extra terms can be interpreted by chosing an alternative integration
contour in the groundstate equations, which surrounds the $1+Y=0$
singularities. Moving back the contour to the real line and picking
up the residues result in the excited state formulas~\cite{Dorey:1996re}. 

In the Lee-Yang model, a single particle can be described by a pair
of complex conjugated roots, $\theta_{1}^{*}=\theta_{2}$ and $\eta_{1}=1=-\eta_{2}$.
See \cite{Dorey:1996re,Bazhanov:1996aq} for details. At large volumes the roots behave as
$\theta_{1}=\theta+i\frac{\pi}{6}+i\delta$, where $\theta$ is the
rapidity of the particle and $\delta$ is exponentially small. The
$SL(2)$ invariant state corresponds to a standing particle ($\theta=0$)
with $\log Y(\theta_{1})=i\pi$. When the volume is decreased, the
root $\theta_{1}$ moves to $i\frac{\pi}{3}$, where it splits into
two, which stay on the $i\frac{\pi}{3}$ line as $\theta_{1}=\theta_{1}^{\pm}=i\frac{\pi}{3}\pm\alpha$,
with $\eta_{1}^{\pm}=\frac{1}{2}=-\eta_{2}^{\pm}$. In the deep UV
$\alpha$ scales as $\alpha=-\log\frac{r}{2}+O(1)$, i.e. one root
is in the kink domain, while, symmetrically, the other one is in the
anti-kink domain. In Appendix~\ref{app.excited} we calculate the $\uvcorr$ coefficient in the small volume expansion
of the energy (\ref{eq:exELY})
\begin{equation}
{\cal R}E_{I}^{TBA}= 2\pi \frac{11}{30}-\frac{\epsilon_{B}}{m^2}r^2+\epsilon_{UV}r^{\frac{24}{5}}+\dots
\end{equation}
The outcome of the calculation
is that in addition to the usual integral term we also have to take
its residues
\begin{equation}
\frac{1}{2}\epsilon_{UV}=-i\sum_{i}\eta_{i}y_L(s_i)-\int\frac{ds}{2\pi}G_{R}\partial\log Y_{R}
\label{e.epsuvres}
\end{equation}
where 
\begin{equation}
G_{R}(s)=\text{\ensuremath{\left\{ \frac{1}{1+Y_{R}(s)}-\frac{1}{1+Y_{0}}\right\} } }y_L\qquad ; \quad y_L(s)=-\tilde{c}2^{-\frac{24}{5}}e^{-\frac{12}{5}s}
\end{equation}
The first term in (\ref{e.epsuvres}) is just the tail of the anti-kink evaluated at the roots $s_i$ up to a factor of $i$. 
The $Y_{R}$ satisfies the right-moving kink equation with the source terms
\begin{equation}
\log Y_{R}=e^{s}+\sum_{i}\eta_{i}\log S(s-s_{i})-\phi\star\log(1+Y_{R}^{-1})\quad;\qquad Y_{R}(s_{i})=-1
\end{equation}
and $Y_{0}=\frac{1}{2}(1-\sqrt{5})$ is its asymptotic value. 
This case is, however, somewhat subtle for numerical analysis as the 
the Y-functions develop zeros on the real line at $\alpha=0.495773177$ (see \cite{Bazhanov:1996aq}) which leads to a singularity in the convolution term.
In order to overcome this difficulty, one can
parametrize the $Y_R$ function as $Y_R(s)=\tanh(\frac{3}{4}(s-\alpha))\,e^{\epsilon_R(s)}$
and solve the modified equation for $\epsilon_R$. See \cite{Bazhanov:1996aq} for details.

A numerical evaluation of (\ref{e.epsuvres}) yields $\uvcorr=-0.000825156 $, which leads to the following result for the mass-gap coefficient from (\ref{e.kappasq}):
\eq
\kappa^2 =  \f{ (2\pi)^{\f{24}{5}}}{2\pi c_2\left(-\f{1}{5}\right)} \uvcorr  = 0.0094185..
\eqx
which agrees well with the exact answer
\eq
\kappa^2= \left( \frac{5^{\frac{3}{4}} \left(\Gamma \left(\frac{2}{3}\right) \Gamma
   \left(\frac{5}{6}\right)\right)^{\frac{12}{5}}}{8\ 2^{\frac{4}{5}} \pi ^{\frac{6}{5}} \Gamma
   \left(\frac{3}{5}\right) \Gamma \left(\frac{4}{5}\right)} \right)^2 = 0.0094184...
\eqx

\section{A Smirnov formula perspective -- resumming\\ the LeClair-Mussardo series}
\label{s.smirnov}

Summarizing the preceding results, the modification of the two external operators $\calO$ in the OPE coefficient which correspond to the particular TBA state is straightforward. 
It is much less clear how to change the inserted operator $\Phi$ in 
\eq
C_{\calO \Phi \calO}
\eqx
from the one generating the massive integrable deformation.
Indeed, as our method involves primarily the computation of the energy correction, it requires \emph{by construction} for $\Phi$ to be equal to the operator appearing in the massive integrable deformation.
But even examining \emph{a-posteriori} the ingredients entering (\ref{e.opely}), we do not see a natural place for modifying the formula to accommodate for changing\footnote{Of course in the Lee-Yang CFT we do not have other operators, but the same kind of formula appears also in the 3-state Potts model -- see Appendix~\ref{s.pottsformula}.} $\Phi$. In this section we will provide a different perspective on the formula (\ref{e.opely}) which has the potential to address this issue, although we do not see its resolution at the current stage.

Returning to the massive integrable deformation, one can rewrite the energy correction related to the OPE coefficient as a 1-point function of the deformation operator. This 1-point function could be obtained from the LeClair-Mussardo series \cite{Leclair:1999ys} which can be resummed, in this case, through a linear ``dressing'' integral equation \cite{Saleur:1999hq}.
In this section we will show that the resulting formulas at the appropriate order in the UV expansion can also be evaluated directly in terms of the CFT integrable data $Y_{R,L}$.

A key motivation for adopting this route is that it has a close structural similarity to Smirnov's approach to ratios of 1-point functions through an underlying fermionic construction \cite{Jimbo:2011bc,Boos:2010qii,Negro:2013wga,Bajnok:2019yik}.
There, the modification of the operator $\Phi$ in $C_{\calO \Phi\calO}$ can be essentially implemented through a modification of the kernel appearing in the linear ``dressing'' integral equation. We leave the investigation of this issue for future work. In the present paper we will still concentrate just on the case of the deforming operator which is universal and does not depend on any underlying microscopic fermionic construction.

The quickest way to reformulate the energy correction in terms of an auxiliary linear dressing equation does not pass through the LeClair-Mussardo formula, but rather starts with rewriting the vacuum expectation value of the trace of the stress tensor $\Theta=T_{00}+T_{11}$ into the combination\footnote{Here we used the perturbed CFT normalization as we would to extract the CFT data.}
\eq
 \frac{1}{m^2}\langle 0 \vert \Theta\vert 0 \rangle_r =  \partial_r E(r) + \f{1}{r} E(r) + \frac{2\epsilon_B}{m^2}
 \label{ThetaVEV}
\eqx
which captures the relevant term in the $r\to 0$ energy expansion.
Here $E(r)=E^{TBA}(r)/m$. The first term can be rewritten as
\eq
\partial_r E(r) = -\partial_r \int \f{d\th}{2\pi} \cosh \th \log(1+Y^{-1})
=\int \f{d\th}{2\pi}  \f{e^\th \partial_r \log Y}{1+Y}
\eqx
where we also used the $\th \to -\th$ symmetry of the problem.
The second term can be rewritten using integration by parts as
\eq
\f{1}{r} E(r) = - \f{1}{r}\int \f{d\th}{2\pi}\f{e^\th \partial_\th \log Y}{1+Y}
\eqx
Putting the two expressions together we get
\eq
\label{e.Hformula}
\partial_r E(r) + \f{1}{r} E(r) = \int \f{d\th}{2\pi}  \f{e^\th H(\th)}{1+Y}
\eqx
with
\eq
H(\th) = \partial_r \log Y - \f{1}{r} \partial_\th \log Y
\eqx
The key property of $H(\th)$ is that it solves a \emph{linear} integral equation with the same kernel as the TBA equation:
\eq
H(\th) = e^{-\th} + \int \f{d\th'}{2\pi} \phi(\th-\th') \f{H(\th')}{1+Y(\th')} 
\label{Htheta}
\eqx
which follows by taking the $r$ and $\th$ derivatives of the standard massive TBA equation (\ref{e.tbamassive}). Denoting the convolution together with the LeClair-Mussardo measure factor $1/(1+Y)$ by $\circ$, we can rewrite it as
\eq
\label{e.Hlinear}
H = e^{-\th} + \phi \circ H
\eqx

The iterative solution for $H(\theta)$ indicates that for large volume\footnote{In (\ref{e.Hlinear}), the dependence on the volume sits implicitly in the $1/(1+Y)$ factor appearing in the $\circ$ convolution.}
$H(\theta)$ is simply $e^{-\theta}$, while for smaller volumes it
is dressed up like 
\begin{equation}
\label{e.Hexpansion}
H(\theta)=e^{-\theta}+e^{-\theta}\phi(\theta-\theta')\circ e^{-\theta'}+\dots=\frac{1}{1-\phi(\theta-\theta')\circ}e^{-\theta'} \equiv (e^{-\theta})^{\mathrm{dr}}
\end{equation}
The expression (\ref{e.Hformula}) can then be rewritten as
\eq
\label{e.Hformulabis}
\partial_r E(r) + \f{1}{r} E(r) = \int \f{d\th}{2\pi}  \f{e^\th H(\th)}{1+Y}  \equiv e^\th \circ (e^{-\theta})^{\mathrm{dr}} \equiv \om_{1,-1}
\eqx
where we used the notation $\omega_{n,m}=e^{n\theta}\circ(e^{m\theta})^{\mathrm{dr}}$ of \cite{Negro:2013wga,Bajnok:2019yik} . Plugging in the UV expansion of the TBA energy (\ref{e.defuvcorr}) we get the following UV expansion of $\om_{1,-1}$ and its link with~$\uvcorr$:
\eq
\om_{1,-1} = -\frac{2 \epsilon_B}{m^2}+\underbrace{\alpha \epsilon_{UV}}_{\omega_{UV}}r^{\alpha -2}+\dots 
\label{e.omuv1}
\eqx
The formulas (\ref{e.Hlinear}) and (\ref{e.Hformulabis}) are indeed a TBA analog of the NLIE formula given in \cite{Jimbo:2011bc}. 
The general approach of \cite{Jimbo:2011bc,Negro:2013wga,Bajnok:2014fca} to ratios of 1-point functions involves exactly equations of the form (\ref{e.Hlinear}), but with the kernel deformed away from the TBA one. We provide further comments on this relation in Appendix~\ref{app.smirnov}.

Yet another interpretation of (\ref{e.Hformulabis}) with the explicit expansion (\ref{e.Hexpansion}), as already indicated above, is that it represents the LeClair-Mussardo series of the trace of the stress tensor as in~\cite{Saleur:1999hq}.

It is actually useful to develop a method to expand directly $\omega_{1,-1}$ for small volumes and express the coefficient $\omega_{UV}$ purely in terms of integrable data of the CFT. This calculation, which we present in Appendix \ref{app.omega} can be generalized for other $\omega$'s, too. Here we summarize the outcome.
The small volume expansion of $\omega_{1,-1}$ in the Scaling Lee-Yang model becomes
\eq
\omega_{1,-1}=\frac{1}{2\sqrt{3}}+r^{\frac{2}{5}}\omega_{UV}+\dots
\label{e.omuv2main}
\eqx
In Appendix~\ref{app.omega}, the quantity $\omega_{UV}$ was rewritten purely in terms of CFT quantities as 
\eq
\omega_{UV}=\int_{-\infty}^{\infty}\frac{ds}{2\pi}g_{R}(s)f_{R}(s) 
\label{e.omegauv}
\eqx
where $g_R$ is the source term governing the deviation of $H$ from its anti-kink counterpart $H_A$ as
\begin{equation}
g_{R}(s)=\tilde{c}\frac{6}{5}\left(\frac{1}{2}\right)^{\frac{7}{5}}e^{-\frac{6}{5}s}\left[\frac{1}{1+Y_{R}(s)}-\frac{1}{1+Y_{0}}\right]+\dots
\label{e.gr}
\end{equation}
This term actually contains the tail of the anti-kink $H_A$ in the kink domain. It turns out to be simply related to the source $G_R$ (see (\ref{e.GKAR})) by
\eq
g_R(s) = -\frac{6}{5} \cdot 2 \cdot G_R(s)
\label{e.gRGRmain}
\eqx
The remaining ingredient of (\ref{e.omegauv}), $f_R$ is the dressed version of $e^s$ in the right domain
\begin{equation}
f_{R}(s)=e^{s}+\frac{\phi(s-s')}{1+Y_{R}(s')}\star f_{R}(s')
\end{equation}
This equation is in fact solved by $f_R(s)=\partial \log Y_{R}(s)$, hence we finally arrive at
\eq
\omega_{UV}=\int_{-\infty}^{\infty}\frac{ds}{2\pi}g_{R}(s)\partial \log Y_{R}(s)
\eqx
Comparing (\ref{e.omuv1}) with (\ref{e.omuv2main}) and using (\ref{e.gRGRmain}), we recover our previous formula (\ref{e.deltaEly}).

\section{Discussion and outlook}

In this paper we initiated a novel approach in which the OPE coefficients of 2D CFTs can be expressed in terms of the integrable characteristics of the CFT, namely in terms of their Y-functions. We focused on the diagonal matrix elements of the primary operators, which define  \emph{integrable} massive perturbations. By expanding the energy levels at small volumes we managed to go beyond the central charge and bulk energy constant terms. We derived a formula for the leading nontrivial  coefficient in the UV expansion from the TBA approach. This is the first term in the perturbative expansion, which goes in  powers of the volume which contain the conformal dimension of the perturbing operator. We expressed this coefficient in terms of the kink (and anti-kink) function -- the solution of the TBA equation in the CFT limit. An essential ingredient of our formula is the tail of the kink function, i.e. the way how it approaches its asymptotic value. The coefficient describing this behaviour is related to the leading nonvanishing  non-local charge~\cite{Bazhanov:1996aq}. 
Actually, what matters for our formula is how the anti-kink function behaves in the kink domain and the kink solution itself. In this sense our formula for the OPE coefficient involves crucially the interaction between the kink and anti-kink domains, in contrast to the conformal weights for which the two domains are clearly separated and independent.

Depending on the state for which we calculate the small volume expansion, our formula can provide 
very valuable quantities. 
If the state is the $SL(2)$ invariant vacuum state, then our formula provides the mass-coupling (mass-gap) relation coefficient. If we choose some other state then our formula can describe CFT 3-point functions, i.e. OPE coefficients. 

We first derived our formula for the ground-state in the Lee-Yang theory, then we also extended it for the simplest excited state, which corresponds to the $SL(2)$ invariant  vacuum state with $h=\bar{h}=0$. These results together determine the mass-coupling relation and the only non-trivial OPE coefficient $C_{\Phi\Phi\Phi}$.
We also generalized our approach to the 3-state Potts minimal model CFT and its two-component TBA system~\cite{Lencses:2014tba}. We investigated both the $SL(2)$ invariant vacuum and the primary field $\sigma\;(\f{1}{15},\f{1}{15})$. Again these results together determine the mass-coupling relation and a non-trivial OPE coefficient $C_{\sg\eps\sg}$. 


The formulas allow for a straightforward extension to the case of arbitrary excited states in a manner consistent with~\cite{Dorey:1996re}. Consequently, one can readily change the operator~$\calO$ in the OPE coefficient $C_{\calO \Phi \calO}$. 
Modifying the operator $\Phi$ remains, however, an outstanding problem, and our original formula does not seem to have a natural place for performing such a modification.   
As a first step in this direction, we rewrote our formula in terms of an auxiliary linear dressing equation, akin to the approach of Smirnov et.al.~\cite{Jimbo:2011bc,Boos:2010qii,Negro:2013wga,Bajnok:2019yik}, where deformations of the kernel in the dressing equation can accommodate for modifying the operator. 
Unfortunately, it is not clear how this works in the generic case which is not related to sine-Gordon theory, like the two component TBA of the (Scaling) Potts model. We leave this investigation for further study.

From a more general perspective,
we believe, that an in-depth study of various integrable descriptions of 2D CFTs, including but not stopping at the level of the spectrum, may lead to interesting insights and a deeper understanding of the scope of integrability.

\bigskip

\noindent{}{\bf Acknowledgments.} RJ would like to thank the Wigner Research Centre for Physics and KITP (within the program \textit{Integrability in String, Field, and Condensed Matter Theory}), while ZB the Jagiellonian University, for hospitality where part of this work was done. This research was supported in part by the National Science Foundation under Grant No. NSF PHY-1748958 and by the NKFIH grant K134946.

\appendix




\section{An analog of the dilogarithm trick for the UV energy correction}
\label{s.dilogenergy}

The subleading energy correction in the kink regime is given by (\ref{e.subleadingtbaE}) which we quote here for convenience:
\eq
\label{e.subleadingtbaErep}
-\int \f{ds}{2\pi} e^s \left[ G_R + \f{\dl Y_R^{-1}}{1+Y_R^{-1}} \right]
\eqx
This expression involves a subleading solution of the massive TBA equation $\dl Y_R^{-1}$. In this section we will show that one can evaluate the above integral without the need to explicitly find $\dl Y_R^{-1}$.
Indeed, let us now follow some initial steps in the dilogarithm trick computation. We differentiate the TBA equation for the kink
\eq
\log Y_R = e^s - \phi \star \log(1+ Y_R^{-1})
\eqx
and integrate by parts the convolution term:
\eq
\int ds' \partial_s \phi(s-s')f(s') = - \int ds' \partial_{s'} \phi(s-s')f(s') = \int ds' \phi(s-s') \partial_{s'} f(s')
\eqx
We then find an expression for $e^s$ 
\eq
e^s = \phi \star \f{\partial Y_R^{-1}}{1+Y_R^{-1}} - \f{\partial Y_R^{-1}}{Y_R^{-1}}
\eqx
Plugging this into (\ref{e.subleadingtbaErep}), we get
\eq
-\int \f{ds ds'}{(2\pi)^2} \phi(s-s') \f{\partial Y_R^{-1}}{1+Y_R^{-1}}(s') \left[ G_R + \f{\dl Y_R^{-1}}{1+Y_R^{-1}} \right](s) +
\int \f{ds}{2\pi} \f{\partial Y_R^{-1}}{Y_R^{-1}} \left[ G_R + \f{\dl Y_R^{-1}}{1+Y_R^{-1}} \right]
\eqx
Performing first the integral over $s$ in the first term and using the linear integral equation (\ref{e.linearizedR}), we get
\eq
-\int \f{ds'}{2\pi} \f{\partial Y_R^{-1}}{1+Y_R^{-1}} \f{\dl Y_R^{-1}}{Y_R^{-1}} + \int \f{ds}{2\pi} \f{\partial Y_R^{-1}}{Y_R^{-1}} \left[ G_R + \f{\dl Y_R^{-1}}{1+Y_R^{-1}} \right]
\eqx
We see that the terms involving $\dl Y_R^{-1}$ cancel out and we are left with a formula involving just the UV limit CFT solutions $Y_{R,L}$.
\eq
 \int \f{ds}{2\pi} G_R \f{\partial Y_R^{-1}}{Y_R^{-1}} 
\equiv -\int \f{ds}{2\pi}  G_R \cdot \partial \log Y_R
\eqx
The computation in the anti-kink region is analogous.

\section{Small volume expansion of the excited state TBA}
\label{app.excited}

In this appendix we explain how excited states TBA equations of the
form 
\begin{equation}
\log Y=r\cosh\theta+\sum_{i}\eta_{i}\log S(\theta-\theta_{i})-\phi\star\log(1+Y^{-1})\quad;\quad Y(\theta_{i})=-1
\end{equation}
can be expanded at small volume. We assume that for $r\to0$ half
of the roots $\{\theta_{i}^{+}\}$ goes to the kink domain and, symmetrically,
the other half $\{\theta_{i}^{-}=-\theta_{i}^{+}\}$ goes to the anti-kink
domain. We introduce the kink equations
\begin{equation}
\log Y_{K}=\frac{r}{2}e^{\theta}+\sum_{i}\eta_{i}\log S(\theta-\theta_{i}^{+})-\phi\star\log(1+Y_{K}^{-1})\quad;\quad Y_{K}(\theta_{i}^{+})=-1
\end{equation}
and define the volume independent right quantities as 
\begin{equation}
Y_{R}(s)=Y_{K}(s-\log\frac{r}{2})\quad;\qquad s_{i}=\theta_{i}^{+}+\log\frac{r}{2}
\end{equation}
We then expand the $Y$-function and the roots around the asymptotic
solution
\[
Y=Y_{\mathrm{as}}+\delta Y=Y_{K}Y_{A}Y_{0}^{-1}+\delta Y\quad;\qquad\theta_{i}=\theta_{i}^{+}+\delta\theta_{i}
\]
Following similar steps we had for the groundstate, we can write the
linearized equations
\begin{equation}
Y_{\mathrm{as}}^{-1}\delta Y=-i\sum_{i}\eta_{i}(\phi(\theta-\theta_{i}^{+})+\phi(\theta+\theta_{i}^{+}))\delta\theta_{i}-\phi\star\left[G-\frac{Y_{\mathrm{as}}^{-1}\delta Y}{1+Y_{\mathrm{as}}}\right]
\end{equation}
where 
\begin{equation}
G=\log\frac{1+Y_{K}^{-1}Y_{A}^{-1}Y_{0}}{1+Y_{K}^{-1}}-\log\frac{1+Y_{0}^{-1}Y_{A}^{-1}Y_{0}}{1+Y_{0}^{-1}}
\end{equation}
The quantization condition is modified as 
\begin{equation}
Y_{A}(\theta_{i}^{+})Y_{0}^{-1}-1=\delta Y(\theta_{i}^{+})+\partial Y_{K}(\theta_{i}^{+})\delta\theta_{i}
\end{equation}
By exploiting that 
\begin{equation}
y_L(s)=\lim_{r\to 0}r^{-\frac{24}{5}}(Y_{A}^{-1}(s-\log\frac{r}{2})Y_{0}-1)=-\tilde{c}2^{-\frac{24}{5}}e^{-\frac{12}{5}s}+\dots
\end{equation}
we can see that the magnitudes of each term is of order of $r^{\frac{24}{5}}$.
We can thus project out the contribution at this order by focusing
on the kink domain and writing only right quantities 
\begin{equation}
Y_{R}^{-1}\delta Y_{R}=-i\sum_{i}\eta_{i}\phi(s-s_{i})\delta s_{i}-\phi\star\left[G_{R}-\frac{Y_{R}^{-1}\delta Y_{R}}{1+Y_{R}}\right]
\end{equation}
and
\begin{equation}-y_L(s_i)=\delta Y_{R}(s_{i})+\partial Y_{R}(s_{i})\delta s_{i}
\end{equation}
where $\delta s_{i}=\delta\theta_{i}$. Fortunately, we do not have
to solve these equations in order to calculate the energy correction
containing $\epsilon_{UV}$. The modification compared to the previous
calculation is the contribution of the extra source terms:
\begin{equation}
-i\sum_{i}\eta_{i}e^{s_i}\delta s_{i}-\int\frac{ds}{2\pi}e^{s}\left[G_{R}-\frac{Y_{R}^{-1}\delta Y_{R}}{1+Y_{R}}\right]
\end{equation}
Here we can again use the adapted dilogarithm trick, namely to differentiate
the $Y_{R}$ kink TBA equation to get $e^{s}$ and plug back to the
energy integral. We then transpose the symmetric kernel and perform
the integral in the second argument. In doing so we exploit the linearized
equations for $Y_{R}^{-1}\delta Y_{R}$ and for $\delta s_{i}$.
On the way we also evaluate the linearized equations and the derivative
of the kink equation at the root positions. Putting everything together
we arrive at the final formula 
\begin{equation}
-i \sum_{i}\eta_{i} y_L(s_i)-\int\frac{ds}{2\pi}G_{R}\,\partial \log Y_{R}
\end{equation}
Observe that the integral term is our generic result, while the discrete
part is merely its residue term, which can be easily seen from 
\begin{equation}
G_{R}=\text{\ensuremath{\left\{ \frac{1}{1+Y_{R}}-\frac{1}{1+Y_{0}}\right\} } } y_L
\end{equation}
by taking the residue in the first term at $s=s_{i}$. 

\section{Comments on the relation with Smirnov's approach}
\label{app.smirnov}

In this section we make a more direct connection with Smirnov's approach, which calculates the vacuum expectation values of exponential vertex operators in terms of $\omega$'s, and evaluate their small volume expansion.

As a first step we need to normalize vacuum expectation values the same way. In the sinh-Gordon theory the ratio of the vacuum expectation
values of the exponential fields can be written in terms of the building
blocks $\omega_{n,m}=e^{n\theta}\circ(e^{m\theta})^{\mathrm{dr}}$.
The simplest of these, $\omega_{1,-1}$, describes the VEV of the
exponential operator, $e^{b\phi}$, in the following way 
\begin{equation}
\frac{\langle 0\vert e^{b\phi}\vert 0\rangle_{r}}{\langle0\vert e^{b\phi}\vert0\rangle_{\infty}}=1+4\sin(\pi p)\omega_{1,-1}\label{eq:shGVEV}
\end{equation}
where the subscript indicates the volume. 
As the $S$-matrix of the sinh-Gordon model coincides with the Lee-Yang
one for $p=-\frac{2}{3}$ we expect that the formula above describes
under this continuation the VEV of $\Phi$. 
We can confirm this by recalling that $\Theta$ is proportional to the perturbing field $\Phi$, so
\eq
\frac{\langle 0 \vert  \Phi  \vert 0  \rangle_{r }}{\langle 0 \vert  \Phi   \vert 0  \rangle_{\infty} } =
\frac{\langle 0 \vert  \Theta  \vert 0  \rangle_{r }}{\langle 0 \vert  \Theta   \vert 0  \rangle_{\infty} }
\eqx
Writing \eqref{ThetaVEV} as
\eq
 \frac{1}{m^2}\langle 0 \vert \Theta\vert 0 \rangle_r =  \om_{1,-1} + \frac{2\epsilon_B}{m^2}
\eqx
and noting that $\omega_{1,-1}=\partial_r E(r) + \f{1}{r} E(r)$ vanishes at large volume,
we see that $\langle 0 \vert  \Theta   \vert 0  \rangle_{\infty} = 2\epsilon_B$, hence putting the two equations together we get
\eq
\frac{\langle 0 \vert  \Phi  \vert 0  \rangle_{r }}{\langle 0 \vert  \Phi   \vert 0  \rangle_{\infty} } = 1+\frac{m^2}{2\epsilon_B} \omega_{1,-1}
\eqx
in agreement with (\ref{eq:shGVEV}) for $p=-\frac{2}{3}$.
For small volumes the
VEV has to approach the finite volume CFT 3-point function as 
\begin{equation}
\lim_{{r}\to0}\langle0\vert\Phi\vert0\rangle_{{r}}=\left (\frac{2\pi}{r}\right)^{2h}C_{\Phi\Phi\Phi}+\dots
\end{equation}


\section{Small volume expansion of the linear dressing formula}
\label{app.omega}

We would like now to express the appropriate UV contribution to $\om_{1,-1}$ in terms of the CFT integrable data. We consider thus the small volume expansion of $\omega_{1,-1}$
\eq
\omega_{1,-1}=\frac{1}{2\sqrt{3}}+r^{\frac{2}{5}}\omega_{UV}+\dots
\label{e.omuv2}
\eqx
and look for an expression for $\omega_{UV}$.

We start by investigating the small volume expansion of $H(\theta)$ defined in \eqref{Htheta}.  Since  $H(\theta)$ is the dressed version of $e^{-\theta}$
it behaves qualitatively similarly, being large in the anti-kink domain.
This large dominating part can be captured by its anti-kink analogue, which satisfies the following equation
\begin{equation}
H_{A}(\theta)=e^{-\theta}+\int_{-\infty}^{\infty}\frac{d\theta'}{2\pi}\frac{\phi(\theta-\theta')H_{A}(\theta')}{1+Y_{A}(\theta')}
\end{equation}
Due to $Y_{A}$ this is a volume-dependent quantity, which can be written in terms of its volume-independent counterpart as 
\begin{equation}
H_{A}\left(\theta\right)=\frac{2}{r}H_{L}(\theta-\log\frac{r}{2})
\end{equation}
where $H_{L}(\theta)$ satisfies the same equation as $H_{A}(\theta)$,
but with $Y_{L}(\theta)$ instead of $Y_{A}(\theta)$. This equation
is actually the same as the equation we have for $-\partial_{\theta}\log Y_{L}(\theta)$,
thus $H_{L}=-Y_{L}^{-1}\partial Y_{L}$. The difference between $H(\theta)$
and $H_{A}(\theta)$ is small for small volumes and satisfies the
linear integral equation: 
\begin{equation}
\delta H(\theta)=H\left(\theta\right)-H_{A}(\theta)=\text{source(\ensuremath{\theta)}+\ensuremath{\phi(\theta-\theta')\circ\delta H(\theta')}}
\end{equation}
where the source term is 
\begin{equation}
\text{source(\ensuremath{\theta)}}=\phi(\theta-\theta') \star {\cal G}(\theta')\quad;\quad{\cal G}(\theta)=H_{A}(\theta)\left[\frac{1}{1+Y(\theta)}-\frac{1}{1+Y_{A}(\theta)}\right]
\end{equation}
To understand the magnitude of ${\cal G}(\theta)$ we point out that
at leading order 
\begin{equation}
\frac{1}{1+Y(\theta)}-\frac{1}{1+Y_{A}(\theta)}=\frac{1}{1+Y_{K}(\theta)}-\frac{1}{1+Y_{0}}+\dots
\end{equation}
which is non-trivial in the kink domain only. We thus need the behaviour
of the anti-kink $H_{A}(\theta)$ in the kink domain, which, being
the interaction between the two domains, is small. Since $H_{L}=-Y_{L}^{-1}\partial Y_{L}$
we can extract its large $\theta$ asymptotics as 
\begin{equation}
Y_{L}(\theta)=Y_{0}(1+\tilde{c}e^{-\frac{6}{5}\theta}+\dots)\to H_{L}(\theta)=\tilde{c}\frac{6}{5}e^{-\frac{6}{5}\theta}+\dots
\end{equation}
which implies, in the variable centered around the kink domain, that
\begin{equation}
H_{A}(\theta-\log\frac{r}{2})=\frac{2}{r}H_{L}(\theta-2\log\frac{r}{2})\sim\tilde{c}\frac{6}{5}\left(\frac{r}{2}\right)^{\frac{7}{5}}e^{-\frac{6}{5}\theta}+\dots
\end{equation}
Thus $\delta H$ starts at order $r^{\frac{7}{5}}$.
We might focus on the first subleading correction by defining the kink
part of the source as 
\begin{equation}
g_{R}(\theta)=\lim_{r\to0}r^{-\frac{7}{5}}{\cal G}(\theta-\log\frac{r}{2})=\tilde{c}\frac{6}{5}\left(\frac{1}{2}\right)^{\frac{7}{5}}e^{-\frac{6}{5}\theta}\left[\frac{1}{1+Y_{R}(\theta)}-\frac{1}{1+Y_{0}}\right]+\dots
\label{gR}
\end{equation}
The corresponding $\delta H_{R}$ satisfies 
\begin{equation}
\delta H_{R}(\theta)=\phi(\theta-\theta') \star \left\{ g_{R}(\theta)+\frac{\delta H_{R}(\theta)}{1+Y_{R}(\theta)}\right\} 
\end{equation}
 and is written purely in terms of CFT right moving quantities.

Eventually we would like to calculate 
\begin{equation}
\omega_{1,-1}=\int_{-\infty}^{\infty}\frac{d\theta}{2\pi}\frac{e^{\theta}(H_{A}(\theta)+\delta H(\theta))}{1+Y(\theta)} =\int_{-\infty}^{\infty}\frac{d\theta}{2\pi}e^{\theta}\left\{ \frac{H_{A}(\theta)}{1+Y_{A}}+ \frac{\delta H(\theta)}{1+Y(\theta)}+{\cal G}(\theta)\right\}  
\end{equation}
The first, leading order term can be calculated from the large $\theta$ asymptotics of $H_{A}(\theta)$.
Indeed, taking the integral equation defining $H_{A}(\theta)$ and
using the large $\theta$ asymptotics of the kernel $\lim_{\theta\to\infty}\phi(\theta)=-2\sqrt{3}e^{-\theta}+\dots.$
we can obtain  
\begin{equation}
\lim_{\theta\to\infty}H_{A}(\theta)=e^{-\theta}\left(1-2\sqrt{3}\int_{-\infty}^{\infty}\frac{d\theta'}{2\pi}\frac{e^{\theta'}H_{A}(\theta')}{1+Y_{A}(\theta')}\right)+\dots
\end{equation}
Since asymptotically $H_{A}(\theta)$ behaves as $H_{A}(\theta)\sim e^{-\frac{6}{5}\theta}$,
 term proportional to $e^{-\theta}$ has to
be cancelled. This implies that the leading, volume-independent term
in $\omega_{1,-1}$ is 
\begin{equation}
\int_{-\infty}^{\infty}\frac{d\theta}{2\pi}\frac{e^{\theta}H_{A}(\theta)}{1+Y_{A}}=\frac{1}{2\sqrt{3}}
\end{equation}
This is related to the bulk energy constant and simply cancels the
$1$ in (\ref{eq:shGVEV}). 

The subleading correction for $\omega_{1,-1}$ has the terms appearing
on the rhs in the integral equation for $\delta H$. Due to the $e^{\theta}$
factor their leading contribution comes from the kink region implying
\begin{equation}
\int_{-\infty}^{\infty}\frac{d\theta}{2\pi}e^{\theta}\left\{ \frac{\delta H(\theta)}{1+Y(\theta)}+{\cal G}(\theta)\right\} =\int_{-\infty}^{\infty}\frac{d\theta}{2\pi}e^{\theta}\left\{ \frac{\delta H_{K}(\theta))}{1+Y_{K}(\theta)}+{\cal G}_{K}(\theta)\right\} +\dots 
\end{equation}
where $\delta H_{K}(\theta)$ and ${\cal G}_{K}(\theta)$ are quantities
in the kink domain, which satisfy the equation 
\begin{equation}
\delta H_{K}(\theta)=\phi(\theta-\theta')\star\left\{ {\cal G}_{K}(\theta')+\frac{\delta H_{K}(\theta))}{1+Y_{K}(\theta)}\right\} 
\end{equation}
In order to extract the volume dependence we make the shift $\theta=s-\log\frac{r}{2}$
and express everything in terms of the CFT right moving quantities:
\eq
r^{\frac{2}{5}}\int_{-\infty}^{\infty}\frac{ds}{2\pi}e^{s}\left\{ \frac{\delta H_{R}(s))}{1+Y_{R}(s)}+g_{R}(s)\right\} 
\eqx
 We can observe that the overall $r$ dependence is $r^{\frac{2}{5}}$
as it is expected from an operator of dimension $h=-\frac{1}{5}$
in the UV limit on the cylinder. Finally, we take such a quantity $f_R$, which satisfies the integral equation
of the form 
\begin{equation}
f_{R}(s)=e^{s}+\frac{\phi(s-s')}{1+Y_{R}(s')}\star f_{R}(s')
\end{equation}
and use the analogue of the dilogarithm trick. Clearly $f_{R}=\partial \log Y_{R}$.
After replacing $e^{s}$ from this integral equation and using
the symmetry of the kernel we obtain 
\begin{equation}
\int_{-\infty}^{\infty}\frac{ds}{2\pi}\text{\ensuremath{\left\{ f_{R}(s)-\phi(s-s')\circ f_{R}(s')\right\} }}\left\{ \frac{\delta H_{R}(s))}{1+Y_{R}(s)}+g_{R}(s)\right\} =\int_{-\infty}^{\infty}\frac{ds}{2\pi}g_{R}(s)f_{R}(s)
\end{equation}
This quantity, which is related to the CFT 3-point functions is written purely in terms of right moving CFT quantities. 

Let us finally relate this quantity to the one appearing in the small volume expansion of the energy. 
The source term $g_R$ governing the deviation of $H$ from its anti-kink counterpart $H_A$ contains the tail of the anti-kink $H_A$ in the kink domain \eqref{gR}. It turns out to be simply related to the source $G_R$ (see (\ref{e.GKAR})) by
\eq
g_R(s) = -\frac{6}{5} \cdot 2 \cdot G_R(s)
\label{e.gRGR}
\eqx
The remaining ingredient $f_R$ is the dressed version of $e^s$ in the right domain, which is in fact given by $f_R(s)=\partial \log Y_{R}(s)$, hence we finally arrive at
\eq
\omega_{UV}=\int_{-\infty}^{\infty}\frac{ds}{2\pi}g_{R}(s)\partial \log Y_{R}(s)
\eqx
Comparing (\ref{e.omuv1}) with (\ref{e.omuv2}) and using (\ref{e.gRGR}), we recover our previous formula (\ref{e.deltaEly}).

\section{Derivation for the 3-state Potts model}
\label{s.pottsformula}
In this Appendix we extend the analysis from the one-component scaling
Lee-Yang TBA system to the two-component scaling Potts model. This
integrable model is obtained by perturbing the D-invariant unitary
${\cal M}_{5,6}$ minimal model with the primary field $\eps$ of dimensions
$(\frac{2}{5},\frac{2}{5})$. Depending on the sign of the perturbation
we can be either in the paramagnetic (single groundstate) or in the
ferromagnetic (triply degenerate groundstate) regime. The scattering
theory in the paramagnetic case has two massive particles, with mass
$m$ which scatter on each other and themselves diagonally \cite{Zamolodchikov:1989cf}. The two-component
TBA equations for the pseudo energies $Y_{1}=e^{\epsilon_{1}}$ and
$Y_{2}=e^{\epsilon_{2}}$ can be written in the compact form
\begin{equation}
\log Y_{i}(\theta)=r\cosh\theta-\phi_{ij}\star\log(1+Y_{j}(\theta)^{-1})
\end{equation}
where the logarithmic derivatives of the scattering matrices $\phi_{ij}(\theta)=-\partial_{\theta}\log S_{ij}(\theta)$
read explicitly as 
\begin{equation}
\phi_{11}=\phi_{22}=-\frac{\sqrt{3}}{1+2\cosh\theta}\quad;\qquad\phi_{12}=\phi_{21}=-\frac{\sqrt{3}}{-1+2\cosh\theta}
\end{equation}
These TBA equations follow from the $A_2$ Y-system relations \cite{Zamolodchikov:1991et}:
\eq
Y_1(\theta -i \frac{\pi}{3}) Y_1(\theta +i \frac{\pi}{3})= 1+Y_2  (\theta) \quad ;\quad Y_2(\theta -i \frac{\pi}{3}) Y_2(\theta +i \frac{\pi}{3})= 1+Y_1  (\theta)
\eqx
The groundstate energy is 
\begin{equation}
E_{0}({\cal R})=-m\sum_{i=1}^{2}\int\frac{d\theta}{2\pi}\cosh\theta\,\log(1+Y_{i}(\theta)^{-1})
\end{equation}

For the groundstate the Y-functions are equal $Y_{1}(\theta)=Y_{2}(\theta)$
and real, thus the only independent TBA equation is the same as the
Lee-Yang one \eqref{e.tbamassive}, while the groundstate energy is doubled. Since the
model is unitary the perturbing operator has a zero
VEV and the leading perturbative correction is a second order one
\begin{equation}
E_{0}({\cal R})=\frac{2\pi}{{\cal R}}\left(-\frac{1}{15}+ c_{2}(2/5)\lambda^{2}\left(\frac{{\cal R}}{2\pi}\right)^{\frac{12}{5}}+\dots\right)
\end{equation}
which corresponds to the integrated two-point function \eqref{c2h} and is equal to twice the formula (\ref{e.cftenergycorrphi}).
Observe that this formula provides the correct central charge $c=\frac{4}{5}$
together with the correct dimension of the perturbing operator $h=\frac{2}{5}$.
By comparing 
\eq
c_{2}(2/5)=\frac{\pi\Gamma\left(\frac{1}{5}\right)\Gamma\left(\frac{2}{5}\right)^{2}}{2\Gamma\left(\frac{3}{5}\right)^{2}\Gamma\left(\frac{4}{5}\right)}
\eqx
and the expansion of the TBA, the mass-coupling
relation $\lambda=\kappa m^{\frac{6}{5}}$ can be established, giving
\eq
\kappa=\frac{\Gamma\left(\frac{3}{10}\right)\left(\Gamma\left(\frac{2}{3}\right)\Gamma\left(\frac{5}{6}\right)\right)^{\frac{6}{5}}\sqrt{\frac{\Gamma\left(-\frac{1}{5}\right)\Gamma\left(\frac{7}{5}\right)}{\Gamma\left(-\frac{2}{5}\right)\Gamma\left(\frac{6}{5}\right)}}}{4\sqrt[5]{2}\pi^{\frac{8}{5}}\Gamma\left(\frac{7}{10}\right)}
\label{e.kappapotts}
\eqx
which agrees with the literature \cite{Fateev:1993av}. 

Interestingly, the perturbing operator has a non-trivial 3-point function
between states of dimensions $(\frac{1}{15},\frac{1}{15})$. In order
to extract this coupling we need to calculate the small volume expansion
of the energy of this excited state. In the paramagnetic phase this
is a genuine excited state with moving particles \cite{Lencses:2014tba}. In the ferromagnetic
phase, however it corresponds to one of the vacua. The ferromagnetic
regime has four kinks which scatter on each other with the same S-matrices
which we had for the paramagnetic regime and give rise to the same
TBA system for the true finite volume groundstate. The finiteness
of the volume lifts the groundstate degeneracy and the twisted vacuum,
corresponding to the state $(\frac{1}{15},\frac{1}{15})$, can be described
by complex twists in the TBA, which implies that $Y_{1}$ and $Y_{2}$
are no longer real. Since they are complex conjugate of each other
we denote $Y_{1}$ by $Y$ while $Y_{2}$ by $\bar{Y}$. The twisted
groundstate TBA equations can be written as~\cite{Lencses:2014tba} 
\begin{equation}
\log Y= \omega +r\cosh\theta-\phi_{11}\star\log(1+Y^{-1})-\phi_{12}\star\log(1+\bar{Y}^{-1})
\end{equation}
together with its complex conjugate, where $\omega =i\frac {2\pi}{3}$.
We now follow analogous steps to the Lee-Yang model to expand the
energy at small volumes and extract the sought for 3-point function. 

In the small volume limit the $Y$ function behaves similarly to the
one in the Lee-Yang model and consists of the kink, anti-kink and
the middle (complex) plateaux parts.
The kink/anti-kink $Y_{K/A}$
function satisfies the $r\cosh\to\frac{r}{2}e^{\pm\theta}$ modified
equations 
\begin{equation}
\log Y_{K}= \omega +\frac{r}{2}e^{\theta}-\phi_{11}\star\log(1+ Y_{K}^{-1})-\phi_{12}\star\log(1+\bar{Y}_{K}^{-1})
\end{equation}
and $Y$ has asymptotic value 
\begin{equation}
Y_{0}=\frac{1-\sqrt{5}}{2}
\end{equation}
which can be obtained from the plateau equation
\begin{equation}
\log Y_{0}=\omega +\frac{1}{3}\log(1+ Y_{0}^{-1})+\frac{2}{3}\log(1+\bar{Y}_{0}^{-1})
\end{equation}
The volume-independent kink functions, which are centered around the
origin are defined as 
\begin{equation}
Y_{R}(s)=Y_{K}(s-\log\frac{r}{2})\quad;\qquad Y_{L}(s)=Y_{A}(s+\log\frac{r}{2})=Y_{R}(-s)
\end{equation}
By inspecting the numerical solution we can observe the following
leading correction to the asymptotic $\theta\to-\infty$ behaviour
\begin{equation}
Y_{R}(s)Y_{0}^{-1}=1+i\tilde{c}e^{\frac{3}{5}s}+\dots 
\quad ;\qquad \tilde{c}=-1.381101886..
\end{equation}
where $\tilde{c}$ is real. The real part of $Y_{R}$ approaches $Y_{0}$ as $e^{\frac{6}{5}\theta}$ and is subleading. 

The asymptotic solution is 
\eq
Y_{\mathrm{as}}=Y_{K}Y_{A}Y_{0}^{-1}
\eqx
and its correction satisfies the equation
\begin{align}
\log Y_{\mathrm{as}}+Y_{\mathrm{as}}^{-1}\delta Y & =\omega+\frac{r}{2}e^{\theta}+\frac{r}{2}e^{-\theta}-\phi_{11}\star\log(1+Y_{\mathrm{as}}^{-1})+\phi_{11}\star\frac{Y_{\mathrm{as}}^{-1}\delta Y}{1+Y_{\mathrm{as}}}\nonumber \\
 & \qquad\qquad-\phi_{12}\star\log(1+\bar{Y}_{\mathrm{as}}^{-1})+\phi_{12}\star\frac{\bar{Y}_{\mathrm{as}}^{-1}\delta\bar{Y}}{1+\bar{Y}_{\mathrm{as}}}
\end{align}
together with is complex conjugate. This can be brought into a two-component
linear integral equation of the form 
\begin{equation}
\left(\begin{array}{c}
\frac{\delta Y}{Y_{\mathrm{as}}}\\
\frac{\delta\bar{Y}}{\bar{Y}_{\mathrm{as}}}
\end{array}\right)=\left(\begin{array}{c}
\mathrm{source}\\
\overline{\mathrm{source}}
\end{array}\right)+\left(\begin{array}{cc}
\phi_{11}\star\frac{1}{1+Y_{\mathrm{as}}} & \phi_{12}\star\frac{1}{1+\bar{Y}_{\mathrm{as}}}\\
\phi_{21}\star\frac{1}{1+Y_{\mathrm{as}}} & \phi_{22}\star\frac{1}{1+\bar{Y}_{\mathrm{as}}}
\end{array}\right)\left(\begin{array}{c}
\frac{\delta Y}{Y_{\mathrm{as}}}\\
\frac{\delta\bar{Y}}{\bar{Y}_{\mathrm{as}}}
\end{array}\right)
\end{equation}
where the convolutions acts on the unknown vectors as well. The source
terms can be written as 
\begin{align}
\mathrm{source} & =-\phi_{11}\star G-\phi_{12}\star\bar{G}
\end{align}
with 
\begin{align}
G & =\log(1+ Y_{K}^{-1}Y_{A}^{-1}Y_{0})-\log(1+ Y_{K}^{-1})-\log(1+ Y_{A}^{-1})+\log(1+ Y_{0}^{-1})\nonumber \\
 & =\log\frac{1+ Y_{K}^{-1}Y_{A}^{-1}Y_{0}}{1+ Y_{K}^{-1}}-\log\frac{1+ Y_{0}^{-1}Y_{A}^{-1}Y_{0}}{1+ Y_{0}^{-1}}
\end{align}
Let us define the volume-independent leading right kink form of the
source as 
\begin{equation}
G_{R}(s)=\lim_{r\to0}r^{-\frac{6}{5}}\mathrm{source}(s-\log\frac{r}{2})
\end{equation}
Using the asymptotics of the anti-kink $Y_{A}$ function in the kink
domain
\begin{equation}
y_L(s)=\lim_{r\to 0}r^{-\frac{6}{5}}(Y_{A}^{-1}(s-\log\frac{r}{2})Y_{0}-1)=-i\tilde{c}\;2^{-\frac{6}{5}}e^{-\frac{3}{5}s}+\dots
\end{equation}
we obtain 
\begin{equation}
G_{R}=\text{\ensuremath{\left\{  \frac{1}{1+Y_{R}}-\frac{1}{1+Y_{0}}\right\} } } y_L
\end{equation}
From the $s\to-\infty$ asymptotics of the kink solution we can
see that 
\begin{equation}
\lim_{s\to-\infty}G_{R}(s)=\frac{ Y_{0}^{-1}}{(1+ Y_{0}^{-1})^{2}}\tilde{c}^{2}(2)^{-\frac{6}{5}}
\end{equation}

We can now proceed to calculate the energy correction. We follow the
calculation in the Lee-Yang model and evaluate
\begin{equation}
\int\frac{d\theta}{2\pi}\cosh\theta\,\log(1+(Y_{\mathrm{as}}^{-1}+\delta Y^{-1}))+\mathrm{cc.}
\end{equation}
By using that 
\begin{equation}
\log(1+ Y_{\mathrm{as}}^{-1})=G+\log(1+ Y_{K}^{-1})+\log(1+ Y_{A}^{-1})-\log(1+ Y_{0}^{-1})
\end{equation}
we can calculate the contributions of the kink/anti-kinks as the central
charge and bulk energy constants. The remaining terms can be evaluated in the
two kink regions
\begin{equation}
\int\frac{ds}{2\pi}\left[G_{R}-\frac{Y_{R}^{-1}\delta Y_{R}}{1+Y_{R}}\right]e^{\theta}+\mathrm{cc}.
\end{equation}
Let us now differentiate the kink TBA 
\begin{equation}
e^{s}=Y_{R}^{-1}\partial Y_{R}-\phi_{11}\star\frac{Y_{R}^{-1}\partial Y_{R}}{1+Y_{R}}-\phi_{12}\star\frac{\bar{Y}_{R}^{-1}\partial\bar{Y}_{R}}{1+\bar{Y}_{R}}
\end{equation}
and plug back to the above form. We then transpose the kernels and
use the linearized integral equations to arrive at 
\eq
\int\frac{ds}{2\pi}G_{R}\,\partial \log Y_{R}+{\rm cc.}
\eqx
This result is written purely in terms of the CFT right moving quantities
and does not require to solve the linearized equation. There is a
similar contribution form the left moving particles.
By adding their contributions we arrive at the 3-point functions
\eq
C_{\sigma \Phi \sigma}\equiv C_{\sg \eps \sg}   =-\frac{1}{\kappa (2\pi)^{\frac{4}{5}}} \left( \int\frac{ds}{2\pi}G_{L}\,\partial \log Y_{L}+\int\frac{ds}{2\pi}G_{R}\,\partial \log Y_{R}+{\rm cc.}\right)
\eqx
where we also used the conventional notation $\eps$ for the operator with $(\f{2}{5}, \f{2}{5})$ in the 3-state Potts minimal model.
A numerical evaluation of this quantity, using (\ref{e.kappapotts}), yields
\eq
C_{\sigma \eps \sigma} = 0.54617761809..
\eqx
while the exact answer is $0.54617761825..$.

Clearly, one can repeat the same steps for the $SL(2)$ vacuum (i.e. TBA with the twist $\om=0$), taking into account different asymptotics of the anti-kink, and recover the mass-gap coefficient (\ref{e.kappapotts}). However, the numerics are essentially identical to the Lee-Yang computation for the reasons explained at the beginning of this section. 


\bibliographystyle{elsarticle-num}
\bibliography{OPETBA}

\begin{thebibliography}{10}
\expandafter\ifx\csname url\endcsname\relax
  \def\url#1{\texttt{#1}}\fi
\expandafter\ifx\csname urlprefix\endcsname\relax\def\urlprefix{URL }\fi
\expandafter\ifx\csname href\endcsname\relax
  \def\href#1#2{#2} \def\path#1{#1}\fi

\bibitem{Belavin:1984vu}
A.~A. Belavin, A.~M. Polyakov, A.~B. Zamolodchikov, {Infinite Conformal
  Symmetry in Two-Dimensional Quantum Field Theory}, Nucl. Phys. B 241 (1984)
  333--380.
\newblock \href {https://doi.org/10.1016/0550-3213(84)90052-X}
  {\path{doi:10.1016/0550-3213(84)90052-X}}.

\bibitem{Poland:2018epd}
D.~Poland, S.~Rychkov, A.~Vichi, {The Conformal Bootstrap: Theory, Numerical
  Techniques, and Applications}, Rev. Mod. Phys. 91 (2019) 015002.
\newblock \href {http://arxiv.org/abs/1805.04405} {\path{arXiv:1805.04405}},
  \href {https://doi.org/10.1103/RevModPhys.91.015002}
  {\path{doi:10.1103/RevModPhys.91.015002}}.

\bibitem{Beisert:2010jr}
N.~Beisert, et~al., {Review of AdS/CFT Integrability: An Overview}, Lett. Math.
  Phys. 99 (2012) 3--32.
\newblock \href {http://arxiv.org/abs/1012.3982} {\path{arXiv:1012.3982}},
  \href {https://doi.org/10.1007/s11005-011-0529-2}
  {\path{doi:10.1007/s11005-011-0529-2}}.

\bibitem{Maldacena:1997re}
J.~M. Maldacena, {The Large N limit of superconformal field theories and
  supergravity}, Adv. Theor. Math. Phys. 2 (1998) 231--252.
\newblock \href {http://arxiv.org/abs/hep-th/9711200}
  {\path{arXiv:hep-th/9711200}}, \href
  {https://doi.org/10.1023/A:1026654312961}
  {\path{doi:10.1023/A:1026654312961}}.

\bibitem{Gromov:2013pga}
N.~Gromov, V.~Kazakov, S.~Leurent, D.~Volin, {Quantum Spectral Curve for Planar
  $\mathcal{N} = 4$ Super-Yang-Mills Theory}, Phys. Rev. Lett. 112~(1) (2014)
  011602.
\newblock \href {http://arxiv.org/abs/1305.1939} {\path{arXiv:1305.1939}},
  \href {https://doi.org/10.1103/PhysRevLett.112.011602}
  {\path{doi:10.1103/PhysRevLett.112.011602}}.

\bibitem{Gromov:2014caa}
N.~Gromov, V.~Kazakov, S.~Leurent, D.~Volin, {Quantum spectral curve for
  arbitrary state/operator in AdS$_{5}$/CFT$_{4}$}, JHEP 09 (2015) 187.
\newblock \href {http://arxiv.org/abs/1405.4857} {\path{arXiv:1405.4857}},
  \href {https://doi.org/10.1007/JHEP09(2015)187}
  {\path{doi:10.1007/JHEP09(2015)187}}.

\bibitem{Gromov:2009tv}
N.~Gromov, V.~Kazakov, P.~Vieira, {Exact Spectrum of Anomalous Dimensions of
  Planar N=4 Supersymmetric Yang-Mills Theory}, Phys. Rev. Lett. 103 (2009)
  131601.
\newblock \href {http://arxiv.org/abs/0901.3753} {\path{arXiv:0901.3753}},
  \href {https://doi.org/10.1103/PhysRevLett.103.131601}
  {\path{doi:10.1103/PhysRevLett.103.131601}}.

\bibitem{Basso:2015zoa}
B.~Basso, S.~Komatsu, P.~Vieira, {Structure Constants and Integrable Bootstrap
  in Planar N=4 SYM Theory} (5 2015).
\newblock \href {http://arxiv.org/abs/1505.06745} {\path{arXiv:1505.06745}}.

\bibitem{Bazhanov:1996aq}
V.~V. Bazhanov, S.~L. Lukyanov, A.~B. Zamolodchikov, {Integrable quantum field
  theories in finite volume: Excited state energies}, Nucl. Phys. B 489 (1997)
  487--531.
\newblock \href {http://arxiv.org/abs/hep-th/9607099}
  {\path{arXiv:hep-th/9607099}}, \href
  {https://doi.org/10.1016/S0550-3213(97)00022-9}
  {\path{doi:10.1016/S0550-3213(97)00022-9}}.

\bibitem{Bajnok:2014fca}
Z.~Bajnok, O.~el~Deeb, P.~A. Pearce, {Finite-Volume Spectra of the Lee-Yang
  Model}, JHEP 04 (2015) 073.
\newblock \href {http://arxiv.org/abs/1412.8494} {\path{arXiv:1412.8494}},
  \href {https://doi.org/10.1007/JHEP04(2015)073}
  {\path{doi:10.1007/JHEP04(2015)073}}.

\bibitem{Zamolodchikov:1989cf}
A.~B. Zamolodchikov, {Thermodynamic Bethe Ansatz in Relativistic Models.
  Scaling Three State Potts and Lee-yang Models}, Nucl. Phys. B 342 (1990)
  695--720.
\newblock \href {https://doi.org/10.1016/0550-3213(90)90333-9}
  {\path{doi:10.1016/0550-3213(90)90333-9}}.

\bibitem{Dorey:1996re}
P.~Dorey, R.~Tateo, {Excited states by analytic continuation of TBA equations},
  Nucl. Phys. B 482 (1996) 639--659.
\newblock \href {http://arxiv.org/abs/hep-th/9607167}
  {\path{arXiv:hep-th/9607167}}, \href
  {https://doi.org/10.1016/S0550-3213(96)00516-0}
  {\path{doi:10.1016/S0550-3213(96)00516-0}}.

\bibitem{Cardy:1989fw}
J.~L. Cardy, G.~Mussardo, {S Matrix of the Yang-Lee Edge Singularity in
  Two-Dimensions}, Phys. Lett. B 225 (1989) 275--278.
\newblock \href {https://doi.org/10.1016/0370-2693(89)90818-6}
  {\path{doi:10.1016/0370-2693(89)90818-6}}.

\bibitem{Dorey:1997rb}
P.~Dorey, R.~Tateo, {Excited states in some simple perturbed conformal field
  theories}, Nucl. Phys. B 515 (1998) 575--623.
\newblock \href {http://arxiv.org/abs/hep-th/9706140}
  {\path{arXiv:hep-th/9706140}}, \href
  {https://doi.org/10.1016/S0550-3213(97)00838-9}
  {\path{doi:10.1016/S0550-3213(97)00838-9}}.

\bibitem{Bajnok:2015eng}
Z.~Bajnok, J.~Balog, K.~Ito, Y.~Satoh, G.~Z. T\'oth, {Exact mass-coupling
  relation for the homogeneous sine-Gordon model}, Phys. Rev. Lett. 116~(18)
  (2016) 181601.
\newblock \href {http://arxiv.org/abs/1512.04673} {\path{arXiv:1512.04673}},
  \href {https://doi.org/10.1103/PhysRevLett.116.181601}
  {\path{doi:10.1103/PhysRevLett.116.181601}}.

\bibitem{Bajnok:2016ocb}
Z.~Bajnok, J.~Balog, K.~Ito, Y.~Satoh, G.~Z. Toth, {On the mass-coupling
  relation of multi-scale quantum integrable models}, JHEP 06 (2016) 071.
\newblock \href {http://arxiv.org/abs/1604.02811} {\path{arXiv:1604.02811}},
  \href {https://doi.org/10.1007/JHEP06(2016)071}
  {\path{doi:10.1007/JHEP06(2016)071}}.

\bibitem{Zamolodchikov:1995xk}
A.~B. Zamolodchikov, {Mass scale in the sine-Gordon model and its reductions},
  Int. J. Mod. Phys. A 10 (1995) 1125--1150.
\newblock \href {https://doi.org/10.1142/S0217751X9500053X}
  {\path{doi:10.1142/S0217751X9500053X}}.

\bibitem{Leclair:1999ys}
A.~Leclair, G.~Mussardo, {Finite temperature correlation functions in
  integrable QFT}, Nucl. Phys. B 552 (1999) 624--642.
\newblock \href {http://arxiv.org/abs/hep-th/9902075}
  {\path{arXiv:hep-th/9902075}}, \href
  {https://doi.org/10.1016/S0550-3213(99)00280-1}
  {\path{doi:10.1016/S0550-3213(99)00280-1}}.

\bibitem{Saleur:1999hq}
H.~Saleur, {A Comment on finite temperature correlations in integrable QFT},
  Nucl. Phys. B 567 (2000) 602--610.
\newblock \href {http://arxiv.org/abs/hep-th/9909019}
  {\path{arXiv:hep-th/9909019}}, \href
  {https://doi.org/10.1016/S0550-3213(99)00665-3}
  {\path{doi:10.1016/S0550-3213(99)00665-3}}.

\bibitem{Jimbo:2011bc}
M.~Jimbo, T.~Miwa, F.~Smirnov, {Fermionic structure in the sine-Gordon model:
  Form factors and null-vectors}, Nucl. Phys. B 852 (2011) 390--440.
\newblock \href {http://arxiv.org/abs/1105.6209} {\path{arXiv:1105.6209}},
  \href {https://doi.org/10.1016/j.nuclphysb.2011.06.016}
  {\path{doi:10.1016/j.nuclphysb.2011.06.016}}.

\bibitem{Boos:2010qii}
H.~Boos, M.~Jimbo, T.~Miwa, F.~Smirnov, {Hidden Grassmann Structure in the XXZ
  Model IV: CFT limit}, Commun. Math. Phys. 299 (2010) 825--866.
\newblock \href {http://arxiv.org/abs/0911.3731} {\path{arXiv:0911.3731}},
  \href {https://doi.org/10.1007/s00220-010-1051-6}
  {\path{doi:10.1007/s00220-010-1051-6}}.

\bibitem{Negro:2013wga}
S.~Negro, F.~Smirnov, {On one-point functions for sinh-Gordon model at finite
  temperature}, Nucl. Phys. B 875 (2013) 166--185.
\newblock \href {http://arxiv.org/abs/1306.1476} {\path{arXiv:1306.1476}},
  \href {https://doi.org/10.1016/j.nuclphysb.2013.06.023}
  {\path{doi:10.1016/j.nuclphysb.2013.06.023}}.

\bibitem{Bajnok:2019yik}
Z.~Bajnok, F.~Smirnov, {Diagonal finite volume matrix elements in the
  sinh-Gordon model}, Nucl. Phys. B 945 (2019) 114664.
\newblock \href {http://arxiv.org/abs/1903.06990} {\path{arXiv:1903.06990}},
  \href {https://doi.org/10.1016/j.nuclphysb.2019.114664}
  {\path{doi:10.1016/j.nuclphysb.2019.114664}}.

\bibitem{Lencses:2014tba}
M.~Lencs\'es, G.~Tak\'acs, {Excited state TBA and renormalized TCSA in the
  scaling Potts model}, JHEP 09 (2014) 052.
\newblock \href {http://arxiv.org/abs/1405.3157} {\path{arXiv:1405.3157}},
  \href {https://doi.org/10.1007/JHEP09(2014)052}
  {\path{doi:10.1007/JHEP09(2014)052}}.

\bibitem{Zamolodchikov:1991et}
A.~B. Zamolodchikov, {On the thermodynamic Bethe ansatz equations for
  reflectionless ADE scattering theories}, Phys. Lett. B 253 (1991) 391--394.
\newblock \href {https://doi.org/10.1016/0370-2693(91)91737-G}
  {\path{doi:10.1016/0370-2693(91)91737-G}}.

\bibitem{Fateev:1993av}
V.~A. Fateev, {The Exact relations between the coupling constants and the
  masses of particles for the integrable perturbed conformal field theories},
  Phys. Lett. B 324 (1994) 45--51.
\newblock \href {https://doi.org/10.1016/0370-2693(94)00078-6}
  {\path{doi:10.1016/0370-2693(94)00078-6}}.

\end{thebibliography}

\end{document}